\documentclass{aastex63}

\usepackage{enumerate}
\usepackage{amssymb} 
\usepackage{amsmath} 

\shorttitle{Exomoons Transiting IPMOs}
\shortauthors{Limbach et al.}

\begin{document}

\title{On the Detection of Exomoons Transiting Isolated Planetary-Mass Objects}

\correspondingauthor{Mary Anne Limbach}
\email{maryannelimbach@gmail.com}

\author[0000-0002-9521-9798]{Mary Anne Limbach}
\affiliation{Department of Physics and Astronomy, Texas A\&M University, 4242 TAMU, College Station, TX 77843-4242 USA}

\author[0000-0003-0489-1528]{Johanna M. Vos}
\affiliation{Department of Astrophysics, American Museum of Natural History, Central Park West at 79th Street, NY 10024, USA}

\author[0000-0002-4265-047X]{Joshua N.\ Winn}
\affiliation{Department of Astrophysical Sciences, Peyton Hall, 4 Ivy Lane, Princeton, NJ 08544, USA}

\author[0000-0002-9831-0984]{Ren\'{e} Heller}
\affiliation{Max Planck Institute for Solar System Research, Justus-von-Liebig-Weg 3, 37077 G\"{o}ttingen, Germany}
\affiliation{Georg-August-Universit{\"a}t G{\"o}ttingen, Institut f{\"u}r Astrophysik, Friedrich-Hund-Platz 1, 37077 G{\"o}ttingen, Germany}

\author[0000-0003-0035-598X]{Jeffrey C. Mason}
\affiliation{Department of Physics, New Mexico Institute of Mining and Technology,  801 Leroy Place, Socorro, NM 87801, USA}

\author[0000-0002-6294-5937]{Adam C. Schneider}
\affiliation{United States Naval Observatory, Flagstaff Station, 10391 West Naval Observatory Rd., Flagstaff, AZ 86005, USA} 
\affil{Department of Physics and Astronomy, George Mason University, MS3F3, 4400 University Drive, Fairfax, VA 22030, USA}

\author[0000-0002-8958-0683]{Fei Dai} 
\affiliation{Division of Geological and Planetary Sciences, California Institute of Technology,
1200 E California Blvd, Pasadena, CA, 91125, USA}

\begin{abstract}
All-sky imaging surveys have identified several dozen isolated planetary-mass objects (IPMOs), far away from any star.
Here, we examine the prospects for detecting transiting moons around these objects.
We expect transiting moons to be common, occurring around 10--15\% of IPMOs,
given that close-orbiting moons have a high geometric transit probability and
are expected to be a common outcome of giant planet formation. 
IPMOs offer an advantage over other directly imaged planets in that high-contrast imaging
is not necessary to detect the photometric transit signal.
For at least 30 ($>\,50$\%) of the currently known IPMOs, observations
of a single transit with the {\it James Webb Space Telescope} would have low enough forecasted noise levels
to allow for the detection of an Io-like or Titan-like moon. Intrinsic variability of
the IPMOs will be an obstacle. Using archival time-series photometry of IPMOs with the {\it Spitzer Space Telescope}
as a proof-of-concept, we found evidence for a fading event of 2MASS\,J1119-1137\,AB that might have been caused by
intrinsic variability, but is also consistent with a single transit of a habitable-zone $1.7\,R_{\Earth}$ exomoon.
Although the interpretation of this particular event is inconclusive,
the characteristics of the data and the
candidate signal suggest that Earth-sized habitable-zone exomoons around IPMOs are detectable with existing instrumentation.
\end{abstract}


\section{Introduction} \label{sec:intro}

Many methods have been suggested to search for the moons of planets outside the
Solar System, which are often called ``exomoons.''
As reviewed by \cite{Heller_2018}, about a dozen signals possibly attributable to exomoons have been described in the literature, based on
gravitational microlensing (\citealt{Bennett_2014}; \citealt{Miyazaki_2018}), signatures in transit spectra (\citealt{Oza_2019}; \citealt{Gebek_2020}), gaps in circumplanetary rings (\citealt{Kenworthy_2015}),
transit-timing variations (TTVs) accompanied by exomoon transits of the host star (\citealt{2018AJ....155...36T,Teachey_2018,2018A&A...617A..49R,Kreidberg_2019,2020AJ....159..142T}), TTVs (\citealt{Fox_2020,2020ApJ...900L..44K}),
direct imaging (\citealt{Lazzoni_2020}), and absorption by gas possibly associated with an orbiting moon (\citealt{Ben_Jaffel_2014}).
Follow-up, confirmation, and further characterization of these exomoon candidates
have proven difficult, making it important to devise more 
methods for detecting exomoons.

Isolated planetary-mass objects (IPMOs) offer another opportunity for exomoon
detection. IPMOs are objects that have the low luminosities
and spectral characteristics expected of giant planets, but can be observed in detail in the absence of a bright host star.
They have also been called free-floating planets, starless planets, or rogue planets.
IPMOs are to be distinguished from the dozen directly-imaged planets
that have been detected via high-contrast imaging
in the vicinity of bright host stars (e.g., \citealt{Bowler_2016}).

Several dozen IPMO candidates have been identified in the literature through their spectral and kinematic signatures of youth and/or low-gravity
(see Table \ref{Table:planets} and references therein).
Depending on their spectral characteristics, IPMOs have been classified as Y, T, or L dwarfs.
IPMOs are objects for which evolutionary models indicate the mass
is less than $13\,M_{\rm Jup}$, making them qualify as planets according
to the deuterium-burning criterion \citep{Spiegel_2011}.
Many of the known IPMOs are young, bright, and well-characterized (e.g. \citealt{2013AJ....145....2F}; \citealt{2013ApJ...777L..20L}; \citealt{2014AJ....147...34S}; \citealt{Gagn__2015}; \citealt{Schneider_2016}; \citealt{2017ApJ...841L...1G}).
Some colder IPMOs with unknown ages have been detected despite their
lower luminosities, by virtue of their
proximity to the Sun.
For example, WISE\,0855-0714 has an estimated mass of 1.5--8\,$M_{\rm Jup}$
and is located 2\,pc from the Sun, giving it an apparent magnitude of 14.0 in the 4.6\,$\mu$m\,{\it WISE}
W2 band \citep{2014ApJ...786L..18L}. There are other Y, T, and L dwarfs 
which may be IPMOs
but for which estimated masses are not available in the literature. IPMOs are most similar to the directly imaged exoplanet population in composition, mass, and age. Directly imaged exoplanets span the same spectral range as IPMOs from Y dwarfs (WD0806-661b; \citealt{luhman2011}) to early L dwarfs ($\beta$ Pic b; \citealt{Lagrange2009, Lagrange2010}). Indeed, some of the objects included in Table \ref{Table:planets}, a list of isolated planetary-mass objects, are companions that are sufficiently separated from their host to allow for variability studies without the use of high-contrast imaging such as the exoplanets Ross 458(AB)c \citep{goldman2010, Scholz2010} and COCONUTS-2b \citep{zhang2021second}.

In this paper, we consider the possibility that IPMOs
have moons similar to the major
moons of Jupiter and Saturn.
The moons of IPMOs might form in one of two ways. If the IPMO
was formerly part of an ordinary planetary system centered on a star,
its moons could have formed in a circumplanetary disk within the larger circumstellar disk,
as is thought to have happened for Jupiter and Saturn \citep{2002AJ....124.3404C, 2003Icar..163..198M, 2010ApJ...714.1052S}. Then, when the planet
was ejected into interstellar space by dynamical interactions, the moons
would have had
a reasonable probability of remaining bound to the planet.
According to calculations by \cite{Rabago_2018} and \cite{Hong_2018}, Io-like moons would have a 55\% \citep{Hong_2018} to 85\% \citep{Rabago_2018} chance of surviving intact.
Alternatively, the IPMO might have formed
in isolation as an unusually low-mass outcome of star-formation processes.
In that case, there would be debate over whether
to call a small companion of an IPMO a ``moon,'' a ``planet,'' or something else.
For simplicity,
we will refer to the secondary body as a moon in either case following a precedent established in the literature \citep{Bennett_2014,Skowron_2014,Miyazaki_2018,2021csss.confE.171T,_vila_2021}. 

In this paper we will show that
young IPMOs are attractive targets for exomoon
searches. In the first place, IPMOs (unlike ordinary planets)
can be observed without the problems associated with the
overwhelming glare of a nearby host star, greatly simplifying exomoon detection
and further characterization.
Second, the currently known examples of young IPMOs
are sufficiently bright
for high signal-to-noise observations with
almost any mid- to large-class ground or space-based telescope.
Because IPMOs emit mainly at infrared wavelengths,
NASA's {\it James Webb Space Telescope (JWST)} is well poised to detect and
study their exomoons. 
These two advantages should make it possible to search
for exomoons around IPMOs in many of the same ways that
astronomers are already searching for exoplanets around nearby stars.
This includes looking for exomoons via direct imaging,
radial-velocity monitoring, astrometric variations,
and transits. Earlier authors have considered applying these methods
to directly-imaged exoplanets orbiting stars
\citep{Lazzoni_2020, Vanderburg_2018, Agol_2015, Cabrera_2007, Heller_2016},
but for IPMOs the observational requirements are more easily
met because high-contrast imaging is unnecessary to detect the IPMO.
Indeed, the gravitational lensing technique has already been used to identify a signal
(MOA-2011-BLG-262Lb) that could have arisen from an exomoon orbiting an IPMO,
although the data do not strongly rule out the possibility that the
signal is from a planet orbiting a low-mass star \citep{Bennett_2014}.

This paper focuses on transit detection because it seemed possible to us that the method can succeed in the near future. As we will argue, existing or near-future instrumentation is capable of detecting exomoons analogous to the solar system moons in systems where it is reasonable to expect the exomoon occurrence rate to be high.
In this sense, our work investigating the prospects for detecting transiting moons around IPMOs is analogous to earlier work on the prospects of finding transiting planets around low-mass stars \citep{2003ApJ...594..533G,2008PASP..120..317N} and brown dwarfs \citep{2013arXiv1304.7248T,Tamburo_2019}.

\section{Expected Exomoon Transit Probabilities, Depths, and Frequencies} \label{sec:stat}

\subsection{Exomoon Transit Probabilities} 

\subsubsection{Transit Probabilities Based on Solar System Moon Statistics}\label{sub:prob}
\begin{figure}
\centering
\includegraphics[width=0.9\textwidth]{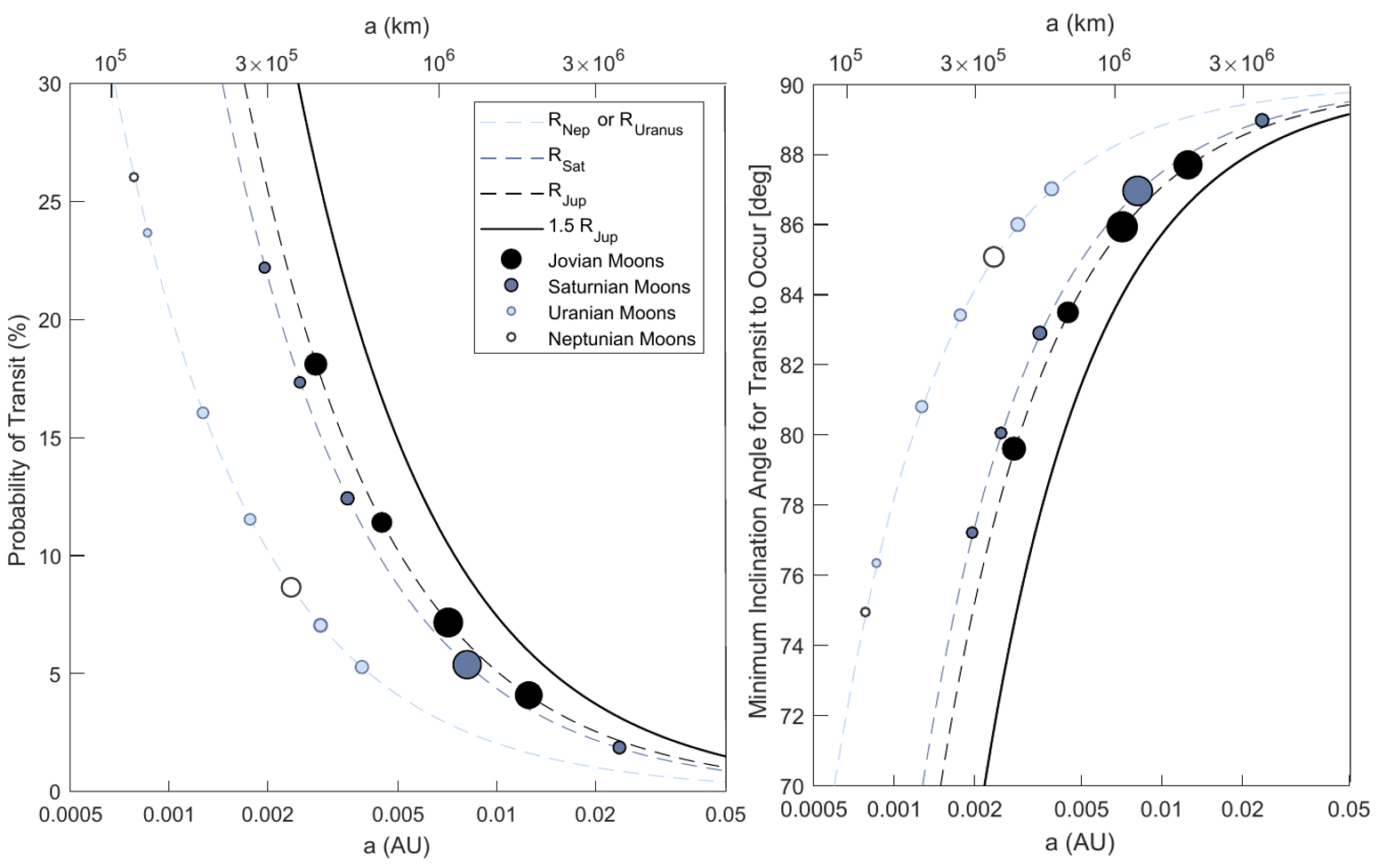}
\caption{{\bf Left:}~Geometric exomoon transit probabilities,
assuming a random viewing direction. The curves correspond to planets
with radii of (from left to right) Uranus/Neptune, Saturn, Jupiter and 1.5
times Jupiter. Circles represent the actual solar system moons with masses
exceeding $4\times 10^{-7}$ of the mass of the host planet.
The median transit probability of these moons is 10\%.
{\bf Right:}~Minimum inclination angle for transits,
versus orbital radius.
The closest-orbiting large moons of the solar system
would transit over an unusually wide range of inclinations.}
\label{ProbDetection}
\end{figure}
We have few observational constraints on the exomoon population
\citep{2015ApJ...806...51H,2018AJ....155...36T}.
However, there is an extensive and growing literature on the formation of moons around gas giants, including $N$-body simulations of moon formation via accretion in circumplanetary disks \citep{2006Natur.441..834C,2012ApJ...753...60O,2015ApJ...806..181H,2015A&A...578A..19H,Miguel_2016,2018MNRAS.475.1347M,2018MNRAS.480.4355C,2020A&A...633A..93R,2020MNRAS.499.1023I,cilibrasi2020nbody} and direct imaging of moon-forming disks \citep{Benisty_2021}. These studies suggest that moon formation around gas giants well separated from a host star is common and that the solar system moons are representative of moon formation around young, accreting giant planets. Simulations show that at least one large moon ($>10^{-6}\,M_{\rm planet}$) forms in $\gtrsim$\,80\% of systems \citep{cilibrasi2020nbody}. Simulations also predict that these moons form in close-in orbits, $\lesssim$\,30 planetary radii, resembling the Galilean moons \citep{Ogihara_2012,2015ApJ...806..181H}. Therefore, lacking any observations of the actual exomoon population, for this work we will assume that IPMOs have moons similar to the moons orbiting the gas giant planets in the solar system. 

We calculate the geometric transit probabilities for the solar system gas giant moons as viewed by an randomly oriented observer outside our solar system. 
Assuming circular orbits and that the secondary companion is much smaller
than the primary object, the geometric transit
probability is $R/a$, the radius of the primary divided by the orbital radius \citep{Winn2010}.
Figure \ref{ProbDetection} (left panel) shows the transit probabilities
for the large moons of the solar system, where ``large'' is
defined as a moon-to-planet
mass ratio exceeding $4\times 10^{-7}$, chosen such that the lowest-mass
moon that qualifies is Neptune's moon Proteus.
The transit probabilities range between about 5\% and 25\%. The median transit
probability of the 16 largest moons is 10\%.
If we restrict the sample to the closest-orbiting large moon around each gas giant planet, the median transit probability is 20\%. Figure \ref{ProbDetection} (right panel) gives the range of inclination angles for transits to occur.
The probability that there is at least one transiting system ($n~\geq~1$) within a
sample of $N$ unrelated primary objects with identical radii ($R$) and with secondary objects at identical orbital distances ($a$) is
\begin{equation}\label{eq:P_geq1_3}
P_{n{\geq}1}(N) = \ 1 - \left(1 - \frac{R}{a}\right)^{\!N}.
\end{equation}

\begin{figure}
\centering
\includegraphics[width=0.9\textwidth]{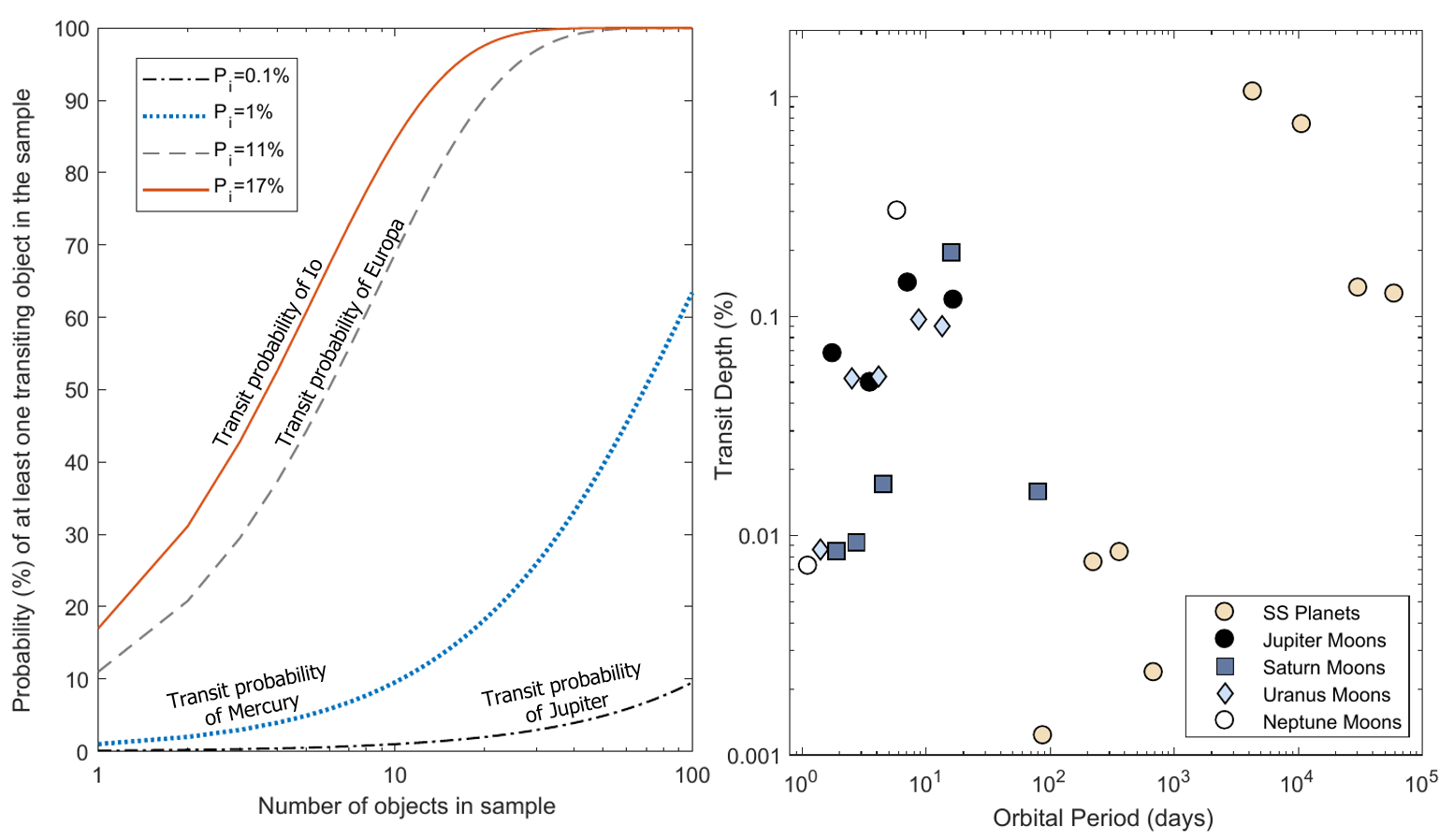}
\caption{{\bf Left:}~Probability that at least one transiting companion exists
in a sample of $N$ unrelated objects with identical secondary companions (Eq.~\ref{eq:P_geq1_3}). Different curves are for different
choices of primary radius and orbital distance.
The blue dotted line is for $R/a=0.01$, similar
to Mercury orbiting the Sun.
The gray dashed curve is for $R/a=0.11$, similar
to Europa orbiting Jupiter.
The red curve is for $R/a=0.17$,
similar Io orbiting Jupiter.
The black dash-dotted curve is for $R/a=0.001$,
similar to Jupiter orbiting the Sun.
{\bf Right:}~Transit depths and orbital periods for selected
solar system moons orbiting solar system planets,
and for solar system planets orbiting the Sun.
}
\label{fig:P_geq1}
\end{figure}

\noindent
Figure~\ref{fig:P_geq1} (left panel) shows $P_{n{\geq}1}(N)$ for four cases.
The black dash-dotted curve is for planets like Jupiter, orbiting
at 5\,AU around Sun-like stars.
The blue dotted line is for planets like Mercury, orbiting
at 0.4\,AU around Sun-like stars.
The gray dashed curve refers to Europa's orbit around Jupiter, and the red curve refers
to Io's orbit around Jupiter.
As an example, in a sample of $N=10$ objects with orbiting secondary companions,
the geometric probability that at least one of the secondaries is transiting
is 10\% for the case of Mercury-Sun analogs,
68\% for Europa-Jupiter analogs, and 85\% for Io-Jupiter analogs.
Figure~\ref{fig:P_geq1} (right panel) shows the fractional loss of light
(transit depth) that would occur during a transit of a moon around a planet,
or a planet around a star, based on solar-system examples.
About 6 solar system moons have transit depths comparable to those of
Neptune and Uranus.

\subsubsection{Transit Probability and the Roche orbital distance}\label{sec:TheoryProbs}
\begin{figure}
\centering
\includegraphics[width=0.95\textwidth]{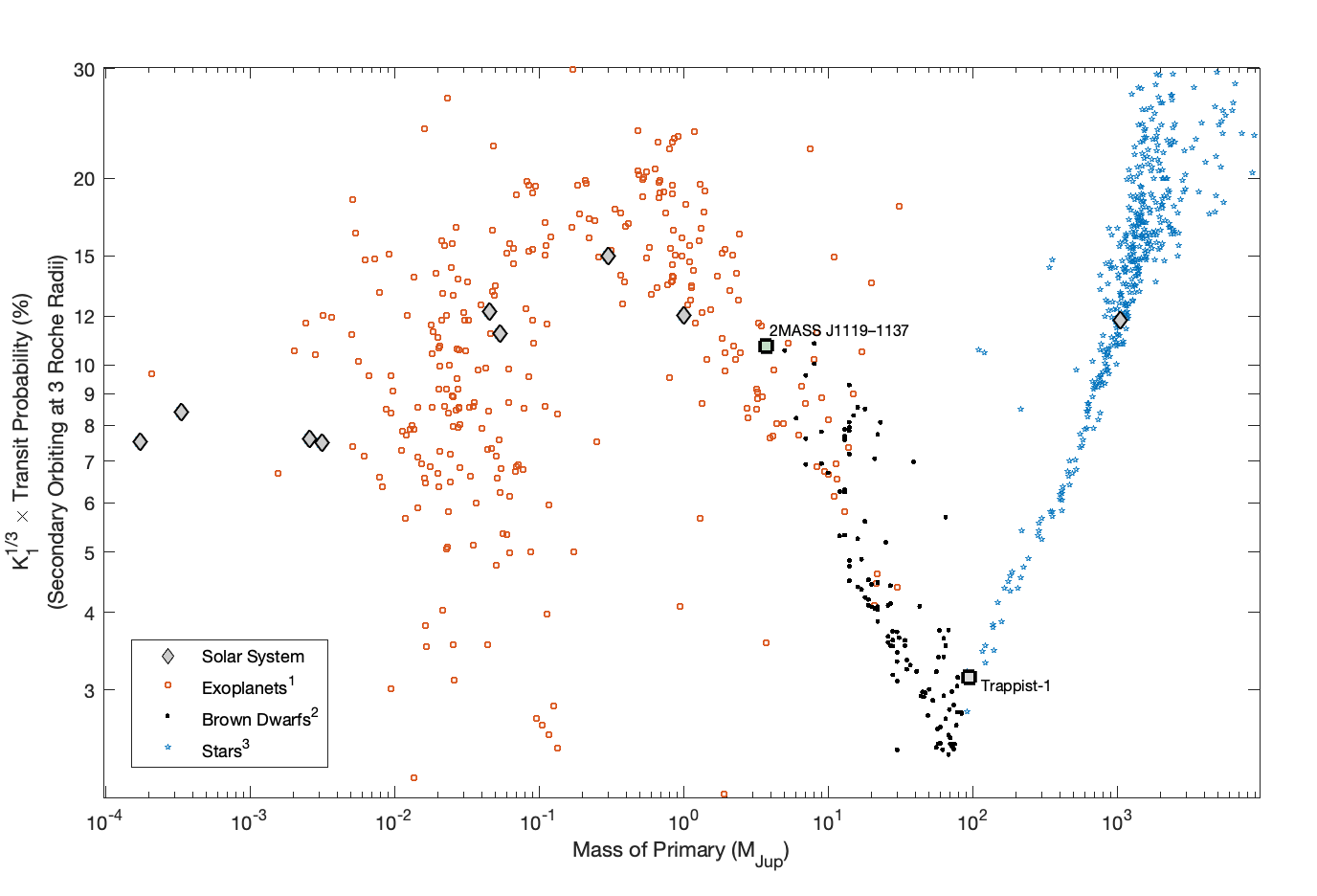}
\caption{Geometric transit probability versus primary mass,
assuming a secondary companion orbiting at 3 Roche radii with a density of $K_1$~g/cm$^3$.
The transit probability has a local maximum for gas giant planets, because of their
large radii and low densities.
The diamonds represent the cases in which the primary object is a planet in the solar system
or the Sun.
Data were taken from [1] the NASA Exoplanet Archive
(only objects with $a>0.1$AU are included in this plot),
[2] \cite{2009AIPC.1094..924G, best2021} and [3] \cite{Parsons_2018, southworth2014debcat}.}
\label{ProbRR}
\end{figure}

Large moons must orbit exterior to the Roche orbital radius.
Here we determine transit probabilities for secondaries that are
located at a fixed multiple of the Roche radius. The Roche radius for a body comprised of an incompressible fluid in a circular orbit is
\begin{equation}\label{RR}
d\simeq2.44\,R_{M}\left({\frac {\rho _{M}}{\rho _{m}}}\right)^{\frac {1}{3}}
\end{equation}
where $R_{M}$ is the primary's radius and $\rho_{M}$ and $\rho_{m}$ are the densities
of the primary and secondary, respectively \citep{Roche_1849}.
We will consider moons for which $a=3d$. This value was chosen because it is close to the mean separation between solar system gas giants and their closest large moons ($a=3.2d$). 
Setting the density to a constant, $K_1$\,g/cm$^3$ for the secondary object, the transit probability at a fixed multiple of the
Roche distance depends only on the properties of the primary object.
Figure \ref{ProbRR} shows the transit probability versus primary mass assuming
a secondary companion orbiting at 3 Roche radii with a density of $K_1$\,g/cm$^3$.
To make this figure, we used data for the masses and densities of
primary objects ranging from small rocky planets to O-type stars. For these calculations, we assume a random orientation, which may not be valid when additional geometric information is available such as the orientation of the planet's orbit.

Starting from the mass of a small rocky planet, as the primary's mass
is increased the transit probability rises (with a large dispersion) until
it reaches a local maximum in the vicinity of gas giant planets.
As the primary mass increases further, the transit probability plunges
because the primary's radius remains roughly constant and the density increases,
moving the Roche radius away from the primary.
The most favorable transit targets in this sense
are Saturn-mass objects.
Specifically, the transit probability at $a=3d$ is 10
times larger for a $0.4\,M_{\rm Jup}$ planet than it is
for a $0.08\,M_\Sun$ star.
Furthermore, young IPMOs and brown dwarfs are often inflated \citep{2015A&A...577A..42B}. For these extremely low density objects, which typically have a radius of up to $1.5\,R_{\rm Jup}$, the transit probability
at a given multiple of the Roche distance is even higher.

\begin{figure}
\centering
\includegraphics[width=0.55\textwidth]{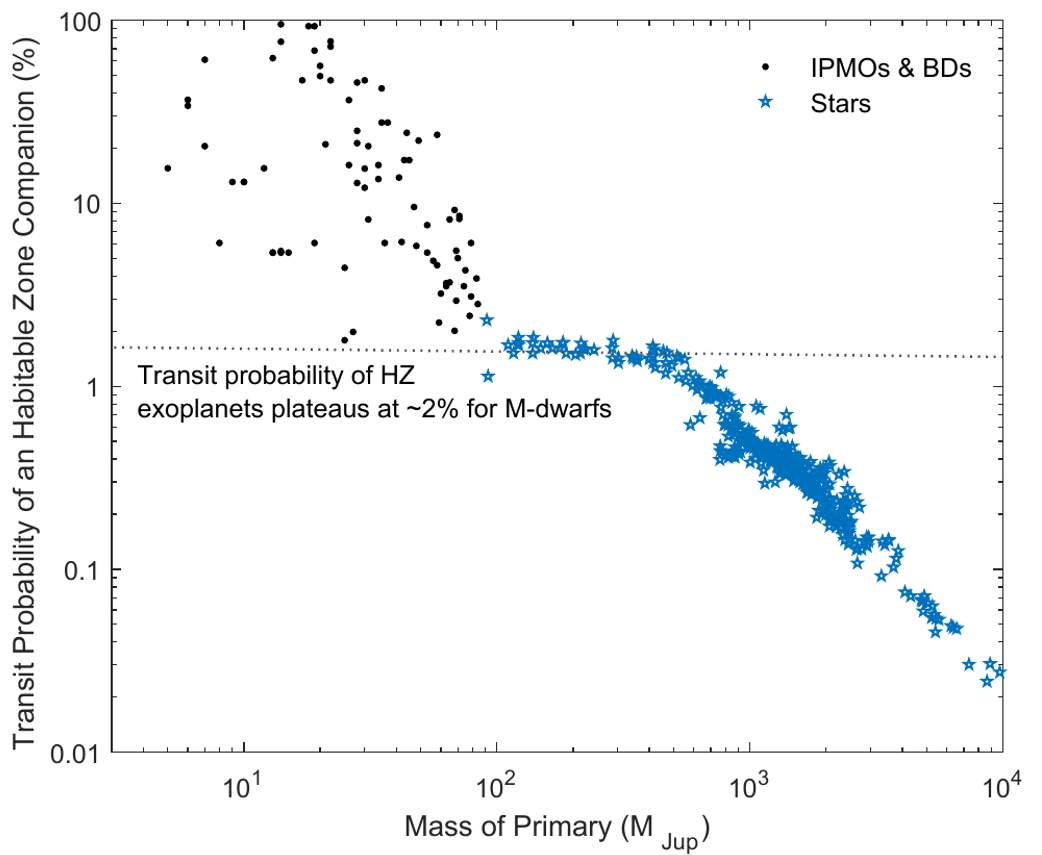}
\caption{Transit probabilities of a companion in the HZ. Here we define the HZ as a companion receiving the same amount of flux as Earth. The transit probabilities of HZ companions is $\sim2\%$ for M-dwarf exoplanets. For more massive stars, the HZ exoplanet transit probability decreases. Transit probabilities for BDs and IPMOs increase with decreasing mass until the primary becomes so cold that a companion can no longer stably orbit in the primary's HZ. Very low-mass BDs and young IPMOs are the most likely objects to host habitable, transiting companions. {\it References:} BDs and IPMOs: \citealt{2009AIPC.1094..924G}; \citealt{best2021}; Stars: \citealt{Parsons_2018}; \citealt{southworth2014debcat}.}
\label{HabProb}
\end{figure}

It is also noteworthy that 3 Roche radii corresponds roughly
to the location
of the habitable zone around young IPMOs with effective temperatures
of approximately 1000\,K.
Figure \ref{HabProb} is similar to Figure~\ref{ProbRR},
but shows the transit probability of a secondary companion
orbiting in the habitable zone instead of at 3 Roche radii.
For this calculation, the habitable zone was
taken to be the orbital distance at which the secondary
receives the same bolometric flux from the primary as Earth does from the Sun.
Starting from a solar mass, as the primary mass is decreased
the HZ transit probability rises until it reaches
a plateau of $2\%$ for M-dwarfs. In the regime of
brown dwarfs and IPMOs, the HZ transit probability resumes
increasing again
until the primary's
luminosity is so low that the HZ is located within the
Roche distance.
In the solar system, the closest-in moons are
at $a/R_{\rm p} \approx 4$ or approximately
2.2 Roche radii (e.g., Enceladus), corresponding to a transit probability of $\approx$\,25\%. Thus, the region of Figure \ref{HabProb} with transit probabilities $\gtrsim$\,25\% may be unrealistic, because
we do not find moons in the solar system in such orbits.

\subsection{Expected Exomoon Masses and Transit Depths}\label{sub:mass}
Most of the solar system moons are thought to have formed via accretion in a circumplanetary disk \citep{2003Icar..163..198M}. The typical moon-to-planet mass fraction for
large moons is observed to be approximately
$2.5\times10^{-4}$, which in theory is the result of the balance between the inflowing material and loss of material through orbital decay \citep{2006Natur.441..834C}. N-body simulations have been used to study solar system moon formation and predict the demographics of exomoon systems  \citep{2012ApJ...753...60O, Heller_2014, Heller_2016, Miguel_2016, 2018MNRAS.475.1347M, 2018MNRAS.480.4355C, 2020A&A...633A..93R, 2020MNRAS.499.1023I, cilibrasi2020nbody}. In particular, \cite{2018MNRAS.480.4355C} used N-body simulations to show that the integrated moon mass of a Jupiter-like planet has a peak between $10^{-4}$ and $10^{-3}$ of the planet's mass, with an upper limit of about 0.1.

If IPMOs host moons with similar mass ratios, then based on these findings, we expect most IPMO systems to have multiple moons with mass  $10^{-4}$ to $10^{-3}\,M_{\rm p}$, transits of which could be detectable with {\it JWST} or
other large infrared-equipped telescopes. For example, for an IPMO with mass $10\,M_{\rm Jup}$, a mass ratio of
a few times $10^{-4}$ corresponds to an Earth-mass moon, and therefore it is reasonable to speculate that Earth-mass moons may be a common outcome of moon formation around IPMOs. Similarly, using the same mass ratio, the analogs of the Galilean satellites around a hypothetical $13\,M_{\rm Jup}$ IPMO
would all be more massive than Mars.

\subsubsection{Exomoon H/He Envelope Capture}
For the preceding calculations, we assumed the mass-radius relationship for exomoons
is the same as observed for solar system moons and rocky exoplanets,
which seems reasonable
but need not be the case in reality.
All of the currently-detected rocky exoplanets have host stars
older than $100$\,Myr. However, most of the known IPMOs are younger than 100\,Myr.
Many of the known short-period rocky exoplanets are thought to have
once had a hydrogen-helium envelope constituting a few percent of the total
mass, based on theoretical interpretations of
the dip at 2\,$R_\oplus$ in the observed radius distribution
(also called the ``radius valley''; \citealt{Lopez_2014}; \citealt{2017AJ....154..109F}; \citealt{Owen_2017}; \citealt{2020MNRAS.498.5030O}; \citealt{10.1093/mnras/stab895};  \citealt{2021MNRAS.503.1526R}).
In these theories, close-orbiting planets below a certain mass threshold
are liable to losing their gaseous envelopes over tens to hundreds of Myrs,
due to high-energy radiation from the primary star or core-powered mass-loss \citep{2021arXiv210503443R}.
Planets as small as Mars ($\gtrsim0.1M_\Earth$) are thought to be sufficiently massive to initially capture H/He during formation (\citealt{1979E&PSL..43...22H,2014P&SS...98..106E,2015A&A...576A..87S,2016ApJ...825...86S}).
\cite{cilibrasi2020nbody} finds that the moons form on a timescale of $10^5$ years. The envelope mass fraction of a $1M_\Earth$ core after being embedded in a disk for $10^5$ years is 1.4\% \citep{2016ApJ...825...86S}, which is a sufficient mass fraction for substantially increasing the radius of the object (\citealt{Rogers_2011}; \citealt{Mordasini_2012}). Although this estimate is given for relevant moon-formation timescales, simulations are for disk conditions (temperature and density) consistent with a planetary nebula at 1\,AU around a sun-like star. Further modeling, beyond the scope of this paper, is needed to understand how envelope capture would differ for the gas temperatures and densities expected in the circumplanetary disk of an IPMO. 
If similar results hold, one might expect large young moons to have H/He envelopes similar to young planets, which would significantly increase their radii and
transit depths.
According to this analogy, a young Earth-mass moon might have a radius
of $2\,R_\Earth$.
Earth-sized moons with H/He envelopes transiting young IPMOs would be less challenging to detect due to exceptionally large transit depths, in the neighborhood of $2\%$.

In summary, based on our analogy with the solar system and on moon formation simulations, it is reasonable to expect that $\gtrsim80\%$ of $10\,M_{\rm Jup}$ IPMOs are hosts of
multiple moons that produce transit depths $\gtrsim$\,0.1\% if the geometry of the system is favorable \citep{cilibrasi2020nbody}.
This is in contrast to what is observed in the exoplanet population. The results of the Kepler mission demonstrated that planets with such large
transit depths and periods shorter than a few days are rare
(\citealt{Beaug__2012}; \citealt{Sanchis_Ojeda_2014}; \citealt{Mazeh_2016}), and hot Jupiters have long been known to have an occurrence of 1\% or lower (\citealt{2012ApJ...753..160W}; \citealt{Zhou_2019}).
In short, there is no known reason to think there is a ``close-in large moon desert'' akin to the ``hot Neptune desert'' and the low occurrence of hot Jupiters.

\subsection{Orbital Periods and Transit Durations}\label{sub:periods}

The orbital periods of exomoons are similar to those of hot Jupiters, and
shorter than those of the solar system planets or typical exoplanets in the
observed sample. Short periods correspond to higher transit probabilities,
and are also helpful for transit detection
by reducing the required amount of observing time to catch multiple transits.
\cite{Heller_2016} calculated that the large moons of the solar system
transit at an average frequency of 0.17/day.
By performing a similar calculation, we find
that the median transit frequency for the {\it closest} large moon orbiting each solar system gas giant planet is 0.55 transits/day (using Io, Tethys, Miranda and Triton) or 0.63 transits/day (if we replace Triton with the closer, lower-mass moon Proteus).
Assuming exomoons have orbital periods similar
to the solar system moons\footnote{This assumption is
roughly consistent with our earlier assumption that moons form at a
typical multiple of the Roche distance, because at a fixed orbital
period, the Roche distance depends only on the mean density of the
moon \citep{Rappaport+2013}.} this implies that on average we need only observe a IPMO for a couple of days in order to achieve sufficient temporal
coverage to detect transiting moons.

\begin{figure}
\centering
\includegraphics[width=0.9\textwidth]{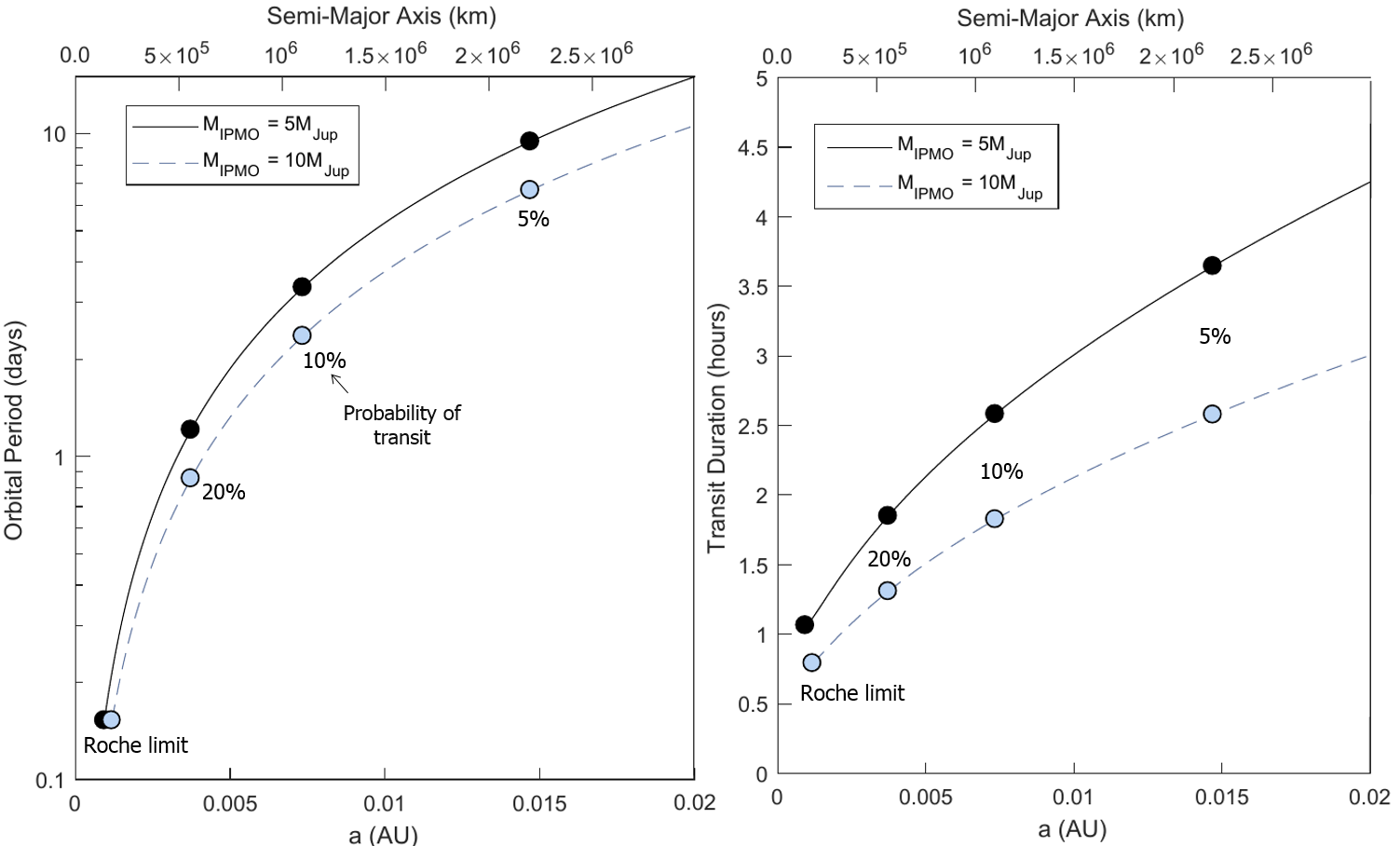}
\caption{Orbital period (left) and transit duration (right, assuming $i=90^\circ$)
versus orbital distance for IPMOs of mass
5\,$M_{\rm Jup}$ (solid black curve)
and 10\,$M_{\rm Jup}$ (dashed curve),
in both cases with radius $1.5\,R_{\rm Jup}$.
The circles and labels indicate the corresponding
transit probabilities and the Roche limit.
Based on the left panel, an observation lasting 20 hours of a 10\,$M_{\rm Jup}$ planet is required to detect all the exomoons with transit probabilities $>\,20\%$ ($a< 0.004$\,AU). Based on the right panel, the exomoons with transit probabilities $>\,20\%$
have transit durations of 1--2 hours.}
\label{ObsTimes}
\end{figure}

Figure \ref{ObsTimes} shows the relationship between exomoon orbital
distance, orbital period (left), transit duration (right) and the probability of exomoon detection. For an edge-on exomoon system, the transit durations are a few hours. Based on our discussion of transit probabilities in Section \ref{sub:prob}, most planets will have a moon with transit probability of 10--20\%, corresponding to an orbital distance
of $\sim10^6$\,km (0.007\,AU). Based on Figure~\ref{ObsTimes}, this implies that if an exomoon is transiting the planet, it will likely be detected within $\sim$50 hours of observations. Thus, if one were interested in surveying planets for transiting exomoons, a minimum of $\sim$50 hours per target would suffice. Experience has shown that detecting transiting exoplanets requires surveys of hundreds or thousands of stars for an interval of at least a few weeks or months. Although wide-field observations of hundreds of IPMOs is not currently possible, there are still good prospects for detecting transiting moons of IPMOs with a few days of observations for each of dozens of objects.

\section{Detectability} \label{sec:detect}

\subsection{Ground- and Space-based Observations}\label{sub:grdspace}

\begin{figure}
\centering
\includegraphics[width=1\textwidth]{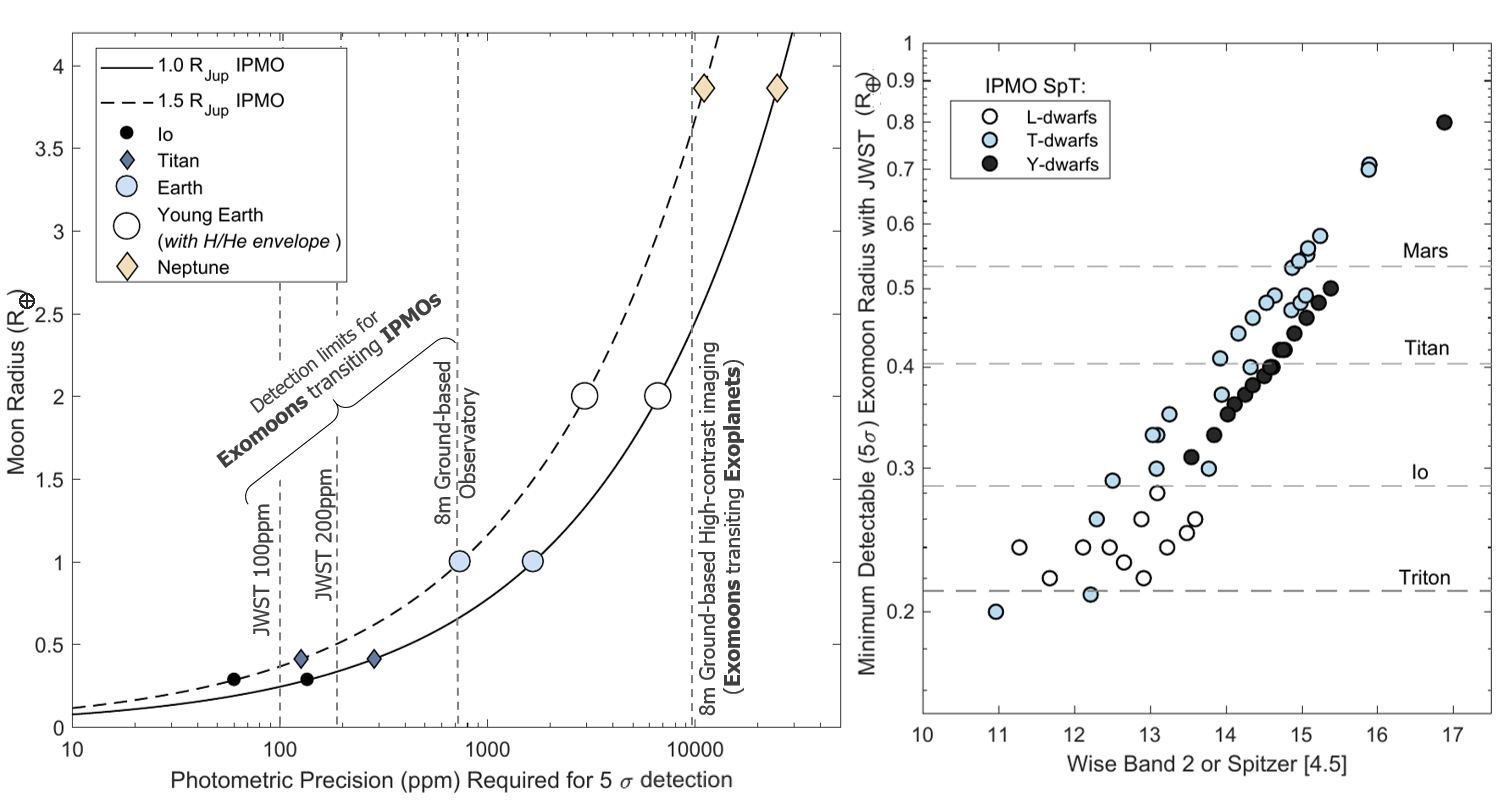}
\caption{{\bf Left:}~Photometric precision required for a 5-$\sigma$ detection of a transiting exomoon,
as a function of the exomoon's radius, assuming the planet's radius is 1.0\,$R_{\rm Jup}$ (solid black line) or $1.5\,R_{\rm Jup}$ (dashed black line). Calculations assume a single 1-hour transit with a JWST NIRSpec photometric precision (JNSPP) of 100ppm or 200ppm, achievable on 17 or 30 of the 57 IPMOs listed in table \ref{Table:planets}, respectively. Vertical dashed gray lines show the photometric sensitivity of various observatories/methods. {\it JWST} can detect analogs of the large moons in the solar system, and ground-based observatories can detect Earth-sized exomoons. Neptune-sized exomoons transiting directly imaged exoplanets are detectable with high-contrast imaging systems.
Earth-sized moons with H/He envelopes (white dots) are more easily detected than purely rocky earths (pale blue dots). {\bf Right:}~Minimum detectable exomoon radius as a function of {\it WISE} Band 2 magnitude (or {\it Spitzer} [4.5] if {\it WISE} magnitude is unavailable) for IPMO calculations given in Table \ref{Table:planets}. Titan-sized moons are detectable around most IPMOs.
}
\label{detectability}
\end{figure}

When trying to detect a transiting moon, the main advantage of IPMOs over other directly-imaged planets is that high-contrast, high-angular-resolution
imaging is not required. 
Using high-contrast imaging techniques, it has proven difficult to achieve a photometric precision better than
1\% level in one hour due to photon noise from the host star and time-variable image artifacts \citep{2020AJ....159..250S}. For example, attempts have been made to search for variability of the HR\,8799 planets,
but the upper limits on any variability are only at the 5--10\% level (\citealt{2016ApJ...820...40A}; \citealt{2021MNRAS.503..743B}).

In contrast, very high photometric precision (200--500\,ppm) within a factor of two of the photon limit
has been achieved for isolated point sources using near-infrared cameras
on ground-based telescopes (\citealt{de_Mooij_2008}, \citealt{Sada_2010}).  The current sample of IPMOs are bright enough
($K \sim 14.5$ mag, $J \sim 16.5$ mag) to support a photon-limited precision of $\sim$0.1\% (1000\,ppm) in one hour with
large (8--10 m) telescope.
For example, a typical near-infrared imager\footnote{Gemini/Flamingos-2 Exposure Time Calculator} on an 8-meter telescope has a photon-limited
Ks-band precision of about 750\,ppm in 1 hour at $K=14.5$ (15 second exposures, 1 hour of total observation time including overheads).  Even if the achievable precision is a factor of two
worse than the photon limit, this would be sufficient
for a 6-$\sigma$ detection of a
1-hour transit of an Earth-sized moon around a Jupiter-sized planet.
Detection of super-Earth and Neptune-sized moons should be possible with 2m-class telescopes.
For example, \cite{Tamburo_2019} showed that a large survey of L and T dwarfs with a small ground-based telescope is likely to
detect at least one transiting planet. They are using this strategy to conduct the Perkins INfrared Exosatellite Survey (PINES) survey in search transiting companions around L and T type brown dwarfs. 

With {\it JWST}, because of the low sky background and stable platform,
it should be possible to improve the photometric precision by an order of magnitude compared to ground-based observatories. It will likely be easier to approach the photon limit with {\it JWST} than with ground-based telescopes, as proved
to be the case for {\it HST} and {\it Spitzer} observations. Figure \ref{detectability} compares the detection limits of various instruments and exomoon detection methods.

Using the JWST/ETC\footnote{jwst.etc.stsci.edu}, we estimated the expected JWST NIRSpec photometric precision (JNSPP) and minimum detectable moon radius for the 57 IPMOs or candidate IPMOs listed Table \ref{Table:planets}. Spectral imaging is preferred over imaging because the dispersed light allows for longer exposures and lower overheads. For transit detection, the total flux signal would initially be more important than the spectral information. 
The spectral information would be useful to check for achromaticity of any flux dips (as would be expected
of transit events), as well as to characterize the planet's own spectrum and intrinsic variability.
This is discussed in more detail in section \ref{sub:var}.

We used the following process to calculate the JNSPP and minimum detectable exomoon radius. We note that the assumptions used in this calculation are approximate for individual objects, but representative of the expected exomoon detection limits for the currently known IPMO population as a whole. However, more precise estimates using models tailored to the exact spectral energy distribution and physical parameters of each IPMO will result in more accurate exomoon detection limits at the individual IPMO level.  

\begin{enumerate}
\item Select a IPMO model from the Sonora 2018 grid \citep{marley2018} based on the spectral type. IPMOs are grouped into five bins/models -- Early L (L0-L4.5): 1700K model, Late L (L5-L9.5): 1200K, Early T (T0-T4.5): 900K, Late T (T5-T9.5): 600K and Early Y: 350K based on \cite{2015ApJ...810..158F}. These five spectral models were uploaded to the JWST/ETC for SNR calculations.
\item The uploaded model is renormalized to the magnitude of each target at 4.5 $\mu$m (or in K-band when a 4.5 $\mu$m flux is unavailable). 
\item Using the JWST/ETC, the SNR in the NIRSpec bright object time series (BOTS) prism (low-res) mode is calculated for each target. This observing mode is used because it allows spectral coverage from $0.6-5\ \mu$m so no adjustment of the bandpass is needed as spectral type/temperature of the IPMO varies. Because of the unprecedented stability of {\it JWST} and NIRSpec's large slit size ($1\farcs6 \times 1\farcs6$), NIRSpec will likely be able to achieve near photon-limited precision. However, if slit losses are a concern or a reference star is needed, simultaneous NIRCam long-wave slitless grism mode + short-wave defocused imaging results in comparable photometric precisions on IPMOs. The detector setup and exposure time is optimized to provide the maximum SNR while avoiding saturation. Note that in a few cases the rapid readout pattern was required to avoid saturation, which increased the overhead and decreased the SNR on the brightest IPMOs. In all cases we require the cadence to be $<1$ minute such that there is sufficient time resolution to observe a transit. The total observation time is set to one hour and SUB2048 array is used for all calculations.
\item The total SNR from the sum of all flux over the entire spectral range ($0.6-5.3\ \mu$m) is then computed and converted to the JWST NIRSpec one-hour photometric precision (JNSPP-1hr) in ppm. This value is given for each target in table \ref{Table:planets}.
\item The minimum detectable moon radius is calculated from the JNSPP for each target using a simple box approximation for the transit. For this calculation we assume that the host is 1.3 $R_{\rm Jup}$ if it is a moving group association member/candidate and presumably young, which typically corresponds to a larger IPMO radius. For field IPMOs, which are typically older, we assume a radius of 1.0 $R_{\rm Jup}$. It is worth noting that for all objects with masses in between those of Saturn and low-mass stars, the radius is generally confined
to the narrow range of 0.9 to 1.5\,$R_{\rm Jup}$. The 5-$\sigma$ minimum detectable moon radius (``Limit'') is given in $R_{\rm Titan}$ and $R_\Earth$ in Table \ref{Table:planets}. 
\end{enumerate}

Based on these calculations, we find that of the 57 IPMOs listed in Table 1, Europa-sized moons are detectable around 11 IPMOs, Ios around 16, Titans or Ganymedes around 33 which is more than half of all known IPMOs, Mars around 49 and Earth-sized moons are detectable around all 57 IPMOs with SNR $> $5-$\sigma$ in 1-hour with JWST/NIRSpec.

While the detection of such small (Galilean-sized) moons should in theory be possible for many IPMOs with {\it JWST}, this level of sensitivity would not be required if 
moon masses are proportional to planet masses ($M_{\rm moons}\approx2\times10^{-4} M_{\rm planet}$; \citealt{2006Natur.441..834C}).
In this scenario, a typical $10\,M_{\rm Jup}$ IPMO would have
moons with sizes in between those of Titan and Earth. As we demonstrated, moons in this mass range should be detectable around almost all known IPMOs.

\startlongtable
\begin{deluxetable*}{ccccccccccc}\label{Table:planets}
\tablenum{1}
\tablecaption{IPMOs and candidate IPMOs}
\tablewidth{0pt}
\tablehead{
\colhead{} & \colhead{} &  \colhead{K} & \colhead{W2\tablenotemark{a}} &  \colhead{Mass\tablenotemark{b}} & \colhead{Assoc.} & \colhead{JNSPP} & \multicolumn{2}{c}{5-$\sigma$ Limit\tablenotemark{e}} & \colhead{}\vspace{-0.2cm}\\
\colhead{Object} & \colhead{SpT} & \colhead{(mag)} &  \colhead{(mag)} & \colhead{($M_{\rm Jup}$)} &  \colhead{Member?} & \colhead{(ppm)} & \colhead{($R_{\rm Titan}$)} & \colhead{($R_\Earth$)} & \colhead{Refs}
}
\startdata
2MASS 02103857-3015313&	L0&	13.50 	 &	12.65 &		13.0$\pm$4.3 & Tuc-Hor & 49 & 0.56 & 0.23 & 10 \\
2MASS J0249-0557c\tablenotemark{c} & L2& 14.78 &  13.59& $11.6\pm1.3$ & $\beta$ Pic & 63 & 0.64 & 0.26 & 16 \\
2MASS 01531463-6744181	&L2	&14.42    &	13.22  &		11.89$\pm$5.36 & Tuc-Hor Cand. 
& 53 & 0.59 & 0.24 & 10\\
2MASS J2208136+292121 & L3 & 14.12 & 12.91 & $12.6\pm0.6$ &  $\beta$ Pic  & 45 & 0.54 & 0.22 & 16 \\
2MASS 03421621-6817321&	L4&	14.54  &	13.48 &	12.4$\pm$6.1 & AB Dor Cand. & 60 & 0.62 & 0.25 & 10 \\
2M1207b\tablenotemark{c} & L6 & 16.93  &... &8$\pm2$ & TW Hya & 165 & 1.04 & 0.42 & 18,19\\
2MASS 22443167+2043433	&L6.5&	14.02 &	12.11 &		10.5$\pm$1.5 & AB Dor & 56 & 0.60 & 0.24 & 10,23,25\\
WISEA J114724.10-204021.3 & L7 & 14.87 & 13.09 &  5--13 & TW Hya Cand. & 74 & 0.69 & 0.28 & 3,4,10 \\
2MASS J11193254-1137466 AB & L7 &  14.75 & 12.88 &  4--8\tablenotemark{d} &  TW Hya Cand. & 66 & 0.66 & 0.26 & 4,5,10 \\
WISE J174102.78-464225.5 & L7 &13.44 & 11.67  & 4--21 & AB Dor Cand. 
& 45 & 0.54 & 0.22 &  7 \\
PSO J318.5338-22.8603 & L7 &14.44 	&	12.46  &		9--15 & $\beta$ Pic & 54 & 0.59 & 0.24 & 8, 10 \\
2MASS 00470038+6803543	&L7	&13.05 	&11.27 &	11.8$\pm$2.6 & AB Dor & 54 & 0.59 & 0.24 &  10,23 \\
HD 203030b\tablenotemark{c} & L7.5 & 16.16 &...  & 11$\pm$4& Field & 91 & 0.59 & 0.24 & 20 \\
2MASS J13243553+6358281 & T2 & 14.06  & 12.29 & 11--12 & AB Dor & 64 & 0.65 & 0.26 & 14,23 \\
SIMP J013656.5+093347 & T2 &  12.6 &  10.96  &	11.7--13.7 & Car-N Cand. 
& 40 & 0.51 & 0.20 & 9,23 \\
ULAS J004757.41+154641.4 &T2 & 16.42 	& 14.86 &  8.3$\pm$0.9 & Argus Cand. & 204 & 1.15 & 0.47 & 17\\
PSO J168.1800-27.2264  & T2.5 & 16.65 	&14.98 &8.0$\pm$0.7 & Argus Cand. & 218 & 1.19 & 0.48 & 17\\
SDSS J152103.24+013142.7 & T3  & 15.57 	&13.94	& 8.5$\pm$0.9 & Argus Cand. & 127 & 0.91 & 0.37 & 17\\
2MASS J00132229-1143006 & T4  & 15.76 	&	14.32&8.1$\pm$0.7 & Argus Cand. & 154 & 1.00 & 0.40 & 17\\
WISEPA J081958.05-033529.0 & T4  & 14.64 & 13.08 &  5.7$\pm$0.5 & $\beta$ Pic Cand. & 83 & 0.73 & 0.30 & 17\\
SDSS J020742.48+000056.2 & T4.5  &16.72  & 15.05 & 7.9$\pm$0.8 & Argus Cand. & 226 & 1.21 & 0.49 & 17\\
WISE J223617.59+510551.9 & T5  & 14.57 & 12.50& 12.1$\pm$1.3 & Car-N Cand. & 79 & 0.71 & 0.29 & 17 \\
SDSS J111010.01+011613.1 & T5.5 & 16.05  & 13.92 &  10--12 & AB Dor & 158 & 1.01 & 0.41 & 6,23 \\
ULAS J154701.84+005320.3 & T5.5 &18.21 & 15.89 & 5.9$\pm$0.9 & Argus Cand. & 470 & 1.75 & 0.71 & 17\\
ULAS J120744.65+133902.7 & T6&  18.67 &15.88& 4.9$\pm$0.8 & Argus Cand. & 467 & 1.74 & 0.70 & 17\\
ULAS J081918.58+210310.4 & T6  & 17.18 &15.24&11.2$\pm$1.2 & AB Dor Cand. & 320 & 1.44 & 0.58 & 17\\
WISEPA J062720.07-111428.8 & T6  &15.51 &13.25 &11.1$\pm$0.9 & AB Dor Cand. & 113 & 0.86 & 0.35 & 17\\
CFBDS J232304.41-015232.3 & T6   &17.23 &15.07& 4.8$\pm$0.7 & $\beta$ Pic Cand. & 290 & 1.37 & 0.55 & 17\\
SDSSp J162414.37+002915.6 & T6  & 15.61 &13.09& 11.0$\pm$0.8 & Car-N Cand. & 105 & 0.83 & 0.33 & 17\\
ULAS J075829.83+222526.7 & T6.5  & 17.87 &15.08&4.8$\pm$0.8 & Argus Cand. & 292 & 1.38 & 0.56 & 17\\
WISE J024124.73-365328.0 & T7 &...& 14.35 & 5.1$\pm$0.4 & Argus Cand. & 197 & 1.13 & 0.46 & 17\\
2MASSI J1553022+153236 & T7  &15.94 &13.03& 12.0$\pm$1.3 & Car-N Cand. & 102 & 0.81 & 0.33 &  17\\
WISE J031624.35+430709.1 & T8  &...&14.64& 4.8$\pm$0.7 & Car-N Cand. & 230 & 1.22 & 0.49 & 17\\
WISEPC J225540.74-311841.8 & T8  & 17.42 &14.16& 2.3$\pm$0.2 & $\beta$ Pic Cand. & 179 & 1.08 & 0.44 & 17 \\
Ross 458(AB)c\tablenotemark{c} & T8 & 16.96  & 13.77  & 6.8--15.9 &  Field & 147 & 0.75 & 0.30 & 1,2 \\
ULAS J130217.21+130851.2 & T8.5  & 18.28 &14.87 &5.6$\pm$0.9 & Car-N Cand. & 260 & 1.30 & 0.53 & 17 \\
WISE J233226.49-432510.6 & T9 & ...& 14.96& 3.8$\pm$0.7 & AB Dor Cand. & 273 & 1.33 & 0.54 & 17\\
UGPS J072227.51-054031.2&	T9	&	17.07	& 	12.21		&	6--9 & Field & 69 & 0.52 & 0.21 & 11 \\
COCONUTS-2b\tablenotemark{c} & T9 &  20.03  & 14.53  & 4.4--7.8 & Field & 217 & 1.19 & 0.48 & 26,27 \\
WISEA J205628.88+145953.6 & Y0  & 20.01 & 13.84 & 8--20 & Field & 175 & 0.82 & 0.33 & 24\\
WISEA J220905.75+271143.6 & Y0 &  ... & 14.77 &  8--19 & Field & 284 & 1.05 & 0.42 & 24\\
WISE J222055.31-362817.4 & Y0 &  21.33 & 14.71 &  8--20 & Field & 275 & 1.03 & 0.42 & 24\\
WISEA J163940.84-684739.4 & Y0 & ... & 13.54  & 5--14 & Field& 151 & 0.76 & 0.31 & 12,15,24\\
WISEA J173835.52+273258.8 & Y0 & 20.58  & 14.50 &  5--14 & Field & 246 & 0.97 & 0.39 & 24\\
WISE J035934.06-540154.6 & Y0 &22.8 & 15.38  & 8--20 & Field & 397 & 1.24 & 0.50 & 12,24\\
WISEA J041022.75+150247.9 & Y0 & 19.91 & 14.11  &8--20 & Field & 201 & 0.88 & 0.36 & 11,24\\
WISEA J114156.67-332635.5 & Y0 &  ...  & 14.61 &  3--8 & Field & 261 & 1.00 & 0.40 & 24\\
WISEA J120604.25+840110.5 & Y0 &  ... &  15.06 &  6--14 & Field & 332 & 1.13 & 0.46 & 24\\
WISEA J082507.37+280548.2 & Y0.5 &  ... & 14.58 & 3--8 & Field & 256 & 0.99 & 0.40 & 24\\
WISEA J035000.31-565830.5 & Y1  &...& 14.75 & 3--8 & Field & 281 & 1.04 & 0.42 & 12,24\\
WISE J064723.23-623235.5 & Y1  & ...  & 15.22  & 5--13 & Field & 363 & 1.18 & 0.48 & 24\\
WD 0806-661b\tablenotemark{c} & Y1 & ... & 16.88 &  7--9 & Field & 1023 & 1.98 & 0.80 & 21,24\\
WISE J154151.65-225024.9 & Y1 & 21.70 & 14.25 &  8--20 & Field & 216 & 0.91 & 0.37 & 11,24\\
WISE J053516.80-750024.9& $\geq$Y1 & ... &  14.90 & 8--20 & Field & 304 & 1.08 & 0.44 & 24\\
WISEA J083011.95+283716.0 & $\geq$Y1 & ... &  16.05 & 4--13 & Field & 591 & 1.51 & 0.61 & 28\\
WISE J085510.83-071442.5& $>$Y2 &  ... &  14.02 & 1.5--8 & Field & 192 & 0.86 & 0.35 & 13,24\\
WISEPA J182831.08+265037.8& $>$Y2 & 23.48 &  14.35  & 3--8 & Field & 228 & 0.94 & 0.38 & 24 
\enddata
\tablenotetext{a}{{\it WISE} band magnitudes. In some cases {\it Spitzer} [4.5] or M-band magnitudes are substituted for W2.}
\tablenotetext{b}{These are model-dependent masses, which are dependent on the assumed age. For young IPMOs, age is based on association with a moving group (i.e. with proper motions, RV and parallax). Some of these IPMOs are candidates of moving groups (rather than bona-fide members). The quoted masses are likely to change as moving group ages are revised in the literature. }
\tablenotetext{c}{Companions that are sufficiently separated from their host to allow for variability monitoring.}
\tablenotetext{d}{Total mass of the binary planet system assuming association with TW Hya; magnitudes are for the unresolved binary.}
\tablenotetext{e}{The ``Limit'' is the estimated 5-$\sigma$ minimum detectable moon radius for each IPMO based on the JNSPP.}
\tablerefs{
[1]	\cite{Burningham_2011}
[2]	\cite{Manjavacas_2019}
[3]	\cite{Schneider_2016}
[4]	Sch18
[5]	Be17
[6]	\cite{Gagn__2015}
[7]	\cite{2014AJ....147...34S}
[8]	\cite{2013ApJ...777L..20L}
[9]	\cite{2017ApJ...841L...1G}
[10] \cite{Faherty_2016}
[11] \cite{Cushing_2011}
[12] \cite{Dupuy_2013}
[13] \cite{2014ApJ...786L..18L}
[14] \cite{Gagn__2018}
[15] \cite{Schneider_2015}
[16] \cite{2018AJ....156...57D}
[17] \cite{2021ApJ...911....7Z}
[18] \cite{2016ApJ...818..176Z}
[19] \cite{2007ApJ...657.1064M}
[20] \cite{2006ApJ...651.1166M}
[21] \cite{2012ApJ...744..135L}
[22] \cite{best2021}
[23] \cite{Kirkpatrick_2021}
[24] \cite{2017ApJ...842..118L}
[25] \cite{2018MNRAS.474.1041V}
[26] \cite{Kirkpatrick_2011}
[27] \cite{zhang2021second}
[28] \cite{2020ApJ...895..145B}.}
\end{deluxetable*}

\begin{figure}
\centering
\includegraphics[width=0.75\textwidth]{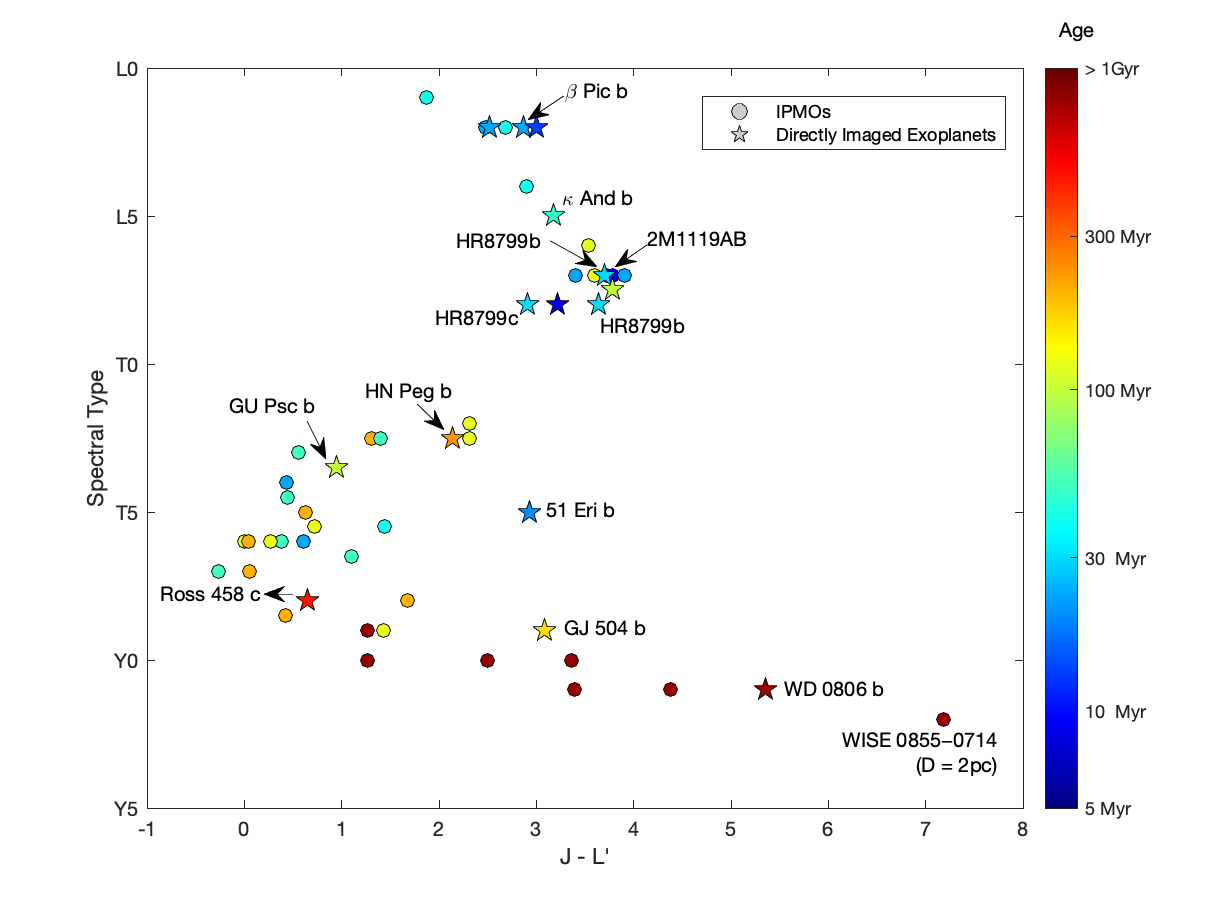}
\caption{Color versus spectral type for directly-imaged exoplanets and IPMOs. All the plotted objects
have estimated masses between 1 and 13 Jupiter masses.
An objects position on this plot is typically set by both age and mass. The hotter L--T objects are generally members of young moving groups, which constrains their masses to be in the planetary regime. The Y dwarfs have not been identified as moving group members, but due to their cold temperatures they lie in the planetary-mass regime regardless of age. 
The directly-imaged exoplanets and IPMOs appear similar, suggesting
they may also have similar satellite systems.
Note the similarities between the planets around HR\,8799 and the IPMO 2MASS\,J1119-1137AB.
In constructing the color, the L' magnitude was used for directly-imaged planets,
whereas the {\it WISE} band 1 or {\it Spitzer} [3.6] band was used for IPMOs when L' was not available.
{\it References:} NASA Exoplanet Archive; \citealt{best2021}.}
\label{IPMOs_DIPlanets}
\end{figure}

Table \ref{Table:planets} provides a list of 57 IPMOs or candidate IPMOs drawn from the literature. We did not include objects that were slightly above the 13$M_{\rm Jup}$ cutoff, even if their error bars (due to an uncertainty in age) indicated that they may be planetary mass (e.g. 2MASS J1324+6358 \citep{Gagn__2018}, a $13.2^{+1.8}_{-1.3}M_{\rm Jup}$ object, was excluded from this list as it is slightly more probable that it is a brown dwarf rather than an IPMO).  As previously noted \citep{2016ApJ...833...96L,Faherty_2016}, IPMOs and directly imaged exoplanets occupy the same ranges of
effective temperature, age, and mass (see Figure \ref{IPMOs_DIPlanets}). It is plausible that these two sets of planetary-mass objects have exomoons with similar properties.
 We prepared this table and figure to convey the typical magnitudes and
other properties of IPMOs and allow for the assessment of exomoon detectability. The IPMOs range in magnitude from 11--16 in the {\it WISE}\,$4.5\,\mu$m band. 
Based on the statistics described in Section \ref{sec:stat},
if exomoons existed around all these systems, and we observed each one for 2 days with {\it JWST},
we would expect several detections.
If no exomoons were detected, we would be able to place meaningful upper limits on exomoon occurrence
that would inform our understanding of IPMOs, exomoon formation, and exomoon survivability during planet ejection (\citealt{Rabago_2018}; \citealt{Hong_2018}). 

For the L dwarfs and early T dwarfs brighter than $K \approx 15$, exomoon transit searches are possible with ground-based K-band observations.
For the planetary-mass Y dwarfs and late-type T dwarfs,
transit searches are probably only feasible by observing further in the infrared range with {\it JWST}. However, because Y dwarfs are typically fainter and have a lower JNSPP, they are generally smaller in radius because they are older. This smaller radius allows for the detection of moons similar in size to younger IPMOs with {\it JWST} despite their faintness.
For example, the Y dwarf WISE 1541-2250 is at a distance of 6\,pc and has a {\it WISE}\,$4.5\,\mu$m band magnitude of 14 despite being more than a Gyr old \citep{Davy_Kirkpatrick_2012} allowing for detection of Titan-sized moons around this IPMO.

\subsection{Substellar Variability}\label{sub:var}

The atmospheres of solar system gas giants, giant exoplanets, and brown dwarfs are characterized by complex chemical processes which often lead to the formation of clouds. The result is rotationally-modulated flux variability produced by cloud features rotating in and out of view. In the case of Jupiter, unresolved observations at $5~\mu$m revealed periodic variability with amplitudes exceeding 20\% \citep{Gelino2000,Ge2019}. 
Both isolated and companion substellar objects are known to exhibit photometric and spectroscopic variability across the full L-Y spectral sequence. This variability probably poses the greatest obstacle to detecting exomoon transits. Large variability surveys from the ground and from space have revealed that photometric variability is common in field brown dwarfs \citep{Buenzli2014, Radigan2014, Metchev2015}. Recent studies have also suggested that variability may be enhanced for the low-gravity brown dwarfs and IPMOs considered here \citep[e.g.][]{Schneider_2018, Vos2019,Vos2020}. Typical infrared amplitudes range from 0.1\% to 5\%, although variations as large as
$25\%$ were observed for the $\approx$\,$13\,M_{\mathrm{Jup}}$ isolated object
2MASS\,J21392676+0220226 \citep{Radigan2012} and the $\approx$\,$20\,M_{\mathrm{Jup}}$
companion VHS\,J1256-1257b \citep{Bowler2020, Zhou2020}.
Y dwarf variability has not yet been studied in great detail, but initial results suggest that they exhibit significant non-sinusoidal variability \citep{Cushing2016, Esplin2016, Leggett2016}.
Therefore, we should expect the amplitude of intrinsic
variations to be on the same order of magnitude (or higher than) the transit signals.

If the periodic variations of the planet remain stable over many cycles,
it may be possible to derive an accurate model for intrinsic
variations and isolate any transit signals. For example, the light curves of the brown dwarfs WISEP\,J190648.47+401106.8 and 2MASS\,J10475385+2124234 
remained stable over many rotations \citep{Gizis2015, Allers2020}.
However, other brown dwarfs have light curves that evolve on rotational timescales, most notably in several L/T transition brown dwarfs such as 2MASS\,J13243553+6358281, SIMP\,J01365662+0933473, and 2MASS\,J21392216+0220185 \citep{Apai2017}. In such cases it will be more difficult to identify a transit. 
Atmospheric dynamical models have shown that light curve evolution is likely very common for brown dwarfs and exoplanets \citep{Tan2021}, but relatively few objects have been monitored longer than $\sim$\,20\,hr. In contrast to evolving cloud-driven variability, the transit signal from an exomoon would have
a consistent depth and duration. Thus, long-term monitoring and searching for repeatable periodic transit signals will be important. Such a search could be done for a handful of brown dwarfs that are bright enough to be observed by the Transiting Exoplanet Survey Satellite (TESS) \citep[e.g.][]{Ricker2015,Apai2021}, but an infrared monitoring campaign would be necessary for a large-scale search for transiting moons.

Another possibility for disentangling intrinsic variations from transits is with spectral time series. There have been many multi-wavelength studies of variability in brown dwarfs and planetary-mass objects. Spectroscopic monitoring using the {\it Hubble Space Telescope (HST)} WFC3 grism camera in particular has revealed the spectral signatures of cloud-driven variability in the atmospheres of both brown dwarfs and planetary-mass objects \citep{Apai2013, Biller2018, Lew2020}. Such signatures sometimes include wavelength-dependent phase shifts due to clouds at different layers in the atmosphere \citep{Buenzli2014,Biller2018}. For example, \citet{Biller2018} observed phase offsets ranging from $200-210^{\circ}$ between simultaneously observed light curves from {\it HST} and the {\it Spitzer Space Telescope} for the free-floating planetary-mass object PSO J318.5-22. Such phase shifts would not occur for
a signal due to a transiting exomoon. Additionally, many studies have characterized the wavelength dependence of the variability amplitude for L-T objects \citep[see][for a compilation]{Manjavacas2019}. The spectral signatures of an exomoon transit are likely to differ from the signatures of clouds that have already been characterized in the literature. 

Despite the additional complications arising from substellar variability for an exomoon search, the existence of variability in a target can have its advantages. By combining rotation periods with high-resolution spectra (from which the projected rotation velocity can be measured), it is possible
to derive a constraint on the viewing inclination for variable brown dwarfs and exoplanets \citep[e.g.][]{Vos2017,Vos2020}. Assuming the moon's orbit is aligned with the planet's rotational axis, this would allow transit surveys to avoid pole-on systems, thereby increasing the average transit probability of the observed
targets \citep[see][for a discussion of the application of this idea to transiting planets]{Beatty2010}.
For example, an Io transit of a Jupiter-sized planet is observable for inclinations between 79.5 and $90^\circ$ (see Figure \ref{ProbDetection}),
giving a geometric transit probability of $\cos(79.5^\circ)\approx 0.18$.
If it were possible to exclude systems with inclinations lower than $60^\circ$ based on prior observations that constrained the IPMO's inclination, the geometric
transit probability would double, becoming $\cos(79.5^\circ)/\cos(60^\circ) \approx 0.36$.
In this respect, $\beta$ Pic\,b offers a particularly favorable geometry for the detection of transiting moons.
If it has a close-in moon system aligned with the star-planet inclination of $89^\circ$ \citep{Kraus_2020}, transits are guaranteed.
For  the  favorable  case  of  an  Earth-mass  moon  with  a  H/He  envelope,  the transit signals would have an amplitude of about 2\%, which might be barely detectable with existing ground-based high-contrast imaging instrumentation.

\subsection{Simulated Exomoon Transits}\label{sub:sim}

\begin{figure}
\centering
\includegraphics[width=.95\textwidth]{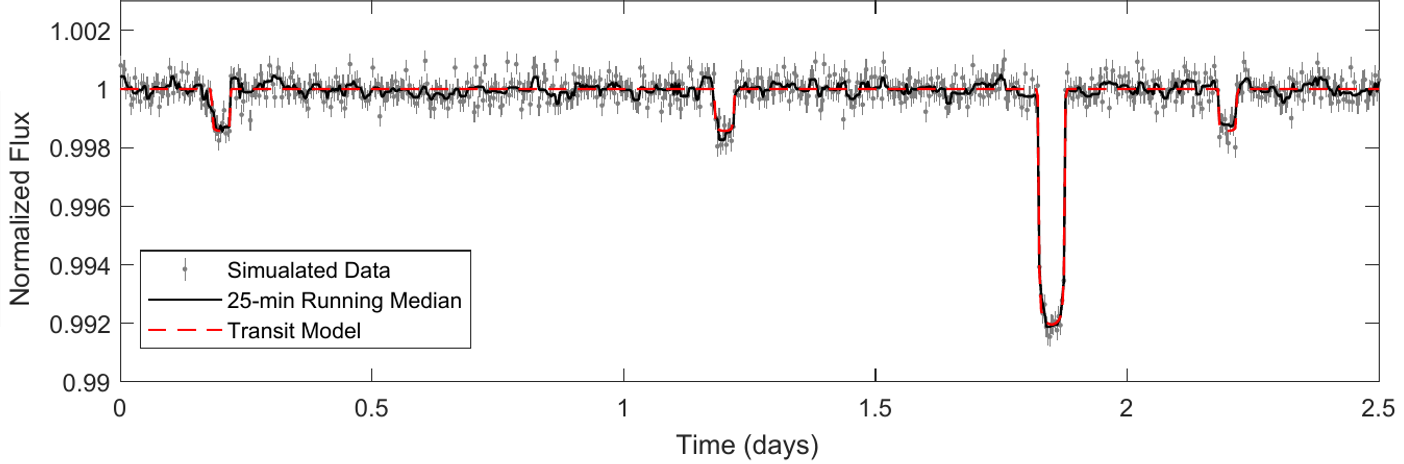}
\caption{Simulated transits observed with {\it JWST}/NIRSpec of an Earth-sized exomoon and a Titan-sized exomoon transiting a 1$R_{\rm Jup}$ IPMO. The transiting Earth- and Titan-sized exomoons are detectable at 97-$\sigma$ and 13-$\sigma$, respectively.}
\label{SimTransit}
\end{figure}
To illustrate the capability of {\it JWST} to detect transiting exomoons around IPMOs,
we simulated light curves of two exomoons transiting a $1\,R_{\rm Jup}$ IPMO using Starry \citep{Luger2019}. 
We generated a transit light curve for an Earth-sized moon with a radius of $1\,R_\Earth$, an orbital period of 3 days and a transit duration of 1.3 hours.
We also generated a light curve for a
Titan-sized moon with a radius of $0.404\,R_{\Earth}$, an orbital period of 1 day and a transit duration of 1 hour in the same synthetic dataset (see Figure \ref{SimTransit}).
 The noise was assumed to follow a Gaussian distribution with a standard deviation
set by the expected sensitivity of the {\it JWST} NIRSpec BOTS mode (a JNSPP-1hr of 100\,ppm or a photometric precision is 350\,ppm per 5 minute bin).
This results in a 97-$\sigma$ detection of each transit of the Earth-like exomoon (duration 1.3 hours, depth 0.8\%)
and a 13-$\sigma$ detection of each transit of the Titan-like exomoon (duration 1 hour, depth 0.13\%).

The goal of this calculation is simply to illustrate that {\it JWST} has sufficient sensitivity to detect exomoons and to demonstrate the expected level of photon-noise in those observations. In reality, observations will likely be limited by variability in the IPMO. The extent to which variability will hinder exomoon detection is currently poorly understood. Section \ref{sec:cand} provides an example showing real IPMO variability in two different photometric bands of {\it Spitzer} light curves. This example illustrates how variability impedes exomoon detection. It also shows how the IPMO variability differs between photometric bands, which demonstrates how we may be able to use spectral information to differentiate between chromatic IPMO variability and (almost) gray exomoons transits. Untangling the spectrally resolved, time-domain variability of IPMOs from exomoon transits in {\it JWST} light curves will require complex IPMO variability modeling coupled with exomoon transit models.


\section{A Fading Event for 2MASS J1119-1137AB} \label{sec:cand}
As a proof-of-concept, we use existing {\it Spitzer} light curves to demonstrate how this exomoon detection method might work and to illustrate some of the challenges of the technique.

\subsection{A Binary IPMO}

2MASS\,J1119-1137AB is a binary IPMO with nearly identical, equal-brightness giant planets separated by $3.9^{+1.9}_{-1.4}$\,AU with an orbital period of $90^{+80}_{-50}$ years (Be17). The source is known to be an equal-brightness binary
based on Keck AO observations in 2016, which gave (Be17)
\begin{equation}
    \log\left(\frac{L_{{\rm bol,\,A}}}{L_\Sun}\right) = -4.73^{+0.27}_{-0.21}~~\textrm{and}~ \log\left(\frac{L_{{\rm bol,\,B}}}{L_\Sun}\right) = -4.74^{+0.27}_{-0.21}.
\end{equation}
For our analysis below, we assumed the two components have equal flux.
The system is a candidate member of the TW Hydrae association (\citealt{Kellogg_2016}; \citealt{2020AJ....159..257B}). If so,
then evolutionary models predict the mass of each planet is $3.7^{+0.9}_{-1.2}\,M_{\rm Jup}$
assuming a system age of $10\pm3$\,Myr (Be17).
If the system is not associated with TW Hya, then it is likely older (10--100\,Myr)
and the estimated mass of each planet is $9.2^{+1.9}_{-2.3} M_{\rm Jup}$ (Be17).
In either case, the masses are consistent with planetary mass sub-stellar objects.
Whether it is associated with TW Hya or not,
the spectrum of 2MASS J1119-1137AB was classified as VL-G (very low gravity), a
signature of youth implying that the binary is young and has low-mass components (Be17).

Based on the possibility that 2MASS\,J1119-1137AB is a relatively isolated member
of the TW Hya association, Be17 argued that the system is a product of the normal star-formation processes, rather than having been ejected via dynamical interactions from a higher-order system. More recent work \citep{2020AJ....159..257B} argues 2MASS J1119-1137AB is older and unassociated with TW Hya, in which case its history might have included
ejection from a star system.
Simulations by \cite{Reipurth_2015} demonstrated that dynamical interactions in triple systems lead naturally to a population of free-floating brown dwarf binaries, however, the simulations did
not include objects lower in mass than $12\,M_{\rm Jup}$.
Detection of the orbital motion of the two IPMOs would lead to constraints
on the masses and thereby shed light on their ages and formation mechanism.

Sch18 observed 2MASS J1119-1137AB with the {\it Spitzer Space Telescope} to measure the rotation periods. We noted an intriguing brightness dip in the published light curve. As described below,
we explored the possibility that this fading
event is due to a transiting moon.

We briefly discuss the WISEA J1147-2040 IPMO light curve (Sch18) in the following analysis as a point of comparison. WISEA J1147-2040 is similar to 2MASS J1119-1137AB in age, distance, spectral type and brightness. It is also a candidate member of TW Hya  with an estimated mass of 5--13 M$_{\rm Jup}$, but it is not known to be a binary \citep{2016ApJ...822L...1S}.

\subsection{Detection of a Fading Event}

The {\it Spitzer Space Telescope} spent 20 hours (12 second exposures) observing 2MASS J1119-1137AB, split equally between the IRAC [3.6] and [4.5] micron bands. An identical observation was performed on WISEA J1147-2040. The IRAC pixel size is 1\farcs2, and
the A and B components of 2MASS J1119-1137 were separated by 0\farcs14; thus they were spatially unresolved.
We used the A+B light curve produced by Sch18.
The resulting light curve is shown in the top panel of Figure \ref{lightcurves}. For comparison, the WISEA J1147-2040 light curve is shown in Appendix \ref{AppB}, Figure \ref{WISEAJ1147}. WISEA J1147-2040 has a much longer rotation period (19.4 hours) compared to 2MASS J1119-1137 (Sch18).

\begin{figure}
\centering
\includegraphics[width=1\textwidth]{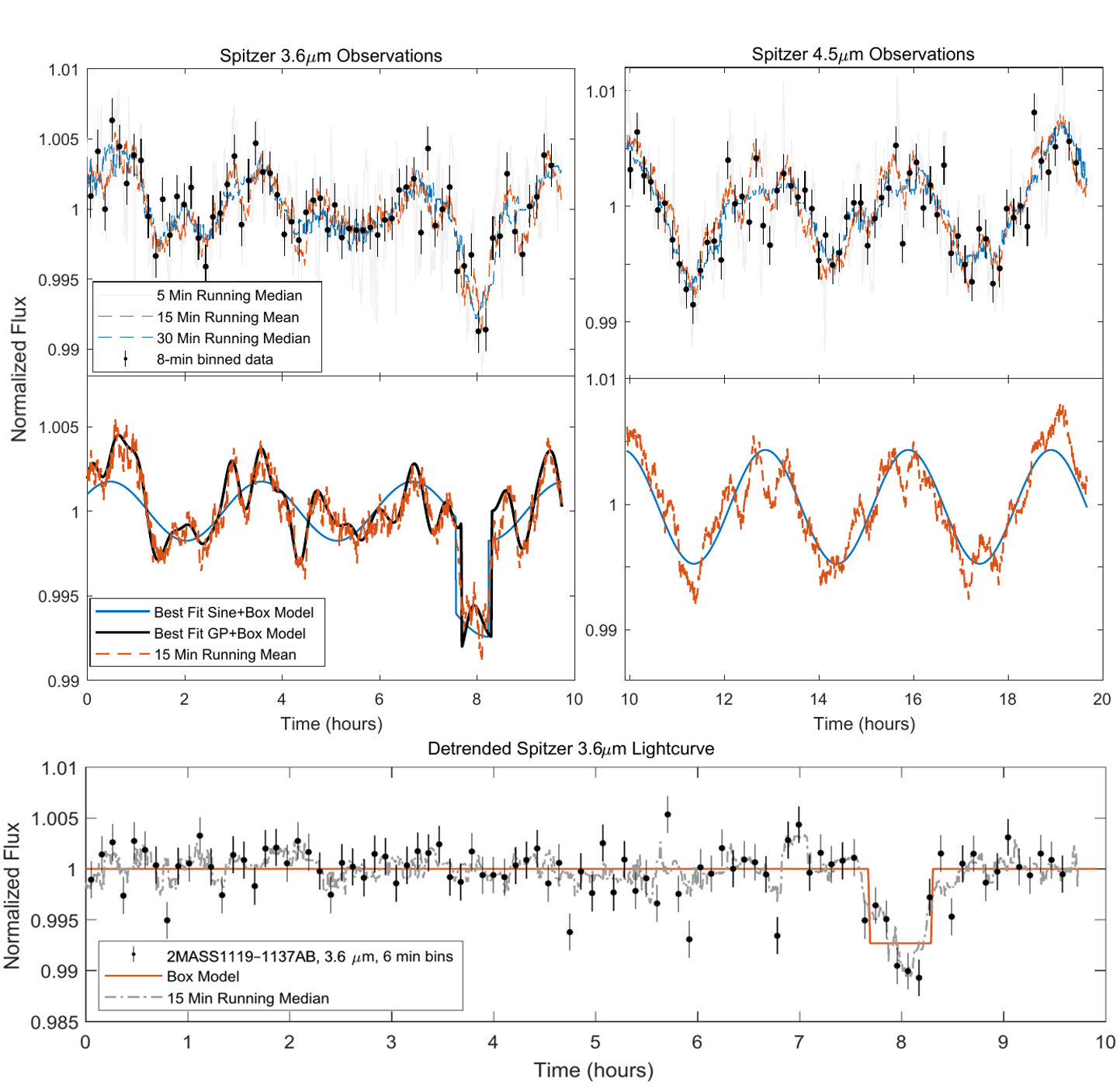}
\caption{{\bf Top:}~{\it Spitzer} light curves of 2MASS J1119-1137AB based on a single 20-hour observation, 10 hours
at $3.6\mu{\rm m}$ (left) and 10 hours
at $4.5\mu{\rm m}$ (right).
Note the fading event at the 8-hour mark
of the $3.6\mu{\rm m}$ light curve.
{\bf Middle:}~Shown in addition to the 15-min running mean of the data (red dashed line) are 
the best fit sine+box model (blue curves)
and the Gaussian Process model of the $3.6\mu m$ data (black curve).
The sine+box model is favored over the sine-only model by $\Delta BIC = 22$. The GP+box model is favored over the GP-only model by a Bayes factor of 3. {\bf Bottom:}~Detrended (with GP-fit) light curve and box fit to data.}
\label{lightcurves}
\end{figure}

To characterize the apparent fading event in the 2MASS J1119-1137 dataset at the
8-hour mark, we fitted the light curve using
two models:
\begin{enumerate}
\item A sinusoidal function, the most commonly used model for IPMO and substellar variability. There are 4 parameters: amplitude, phase, period, and mean flux.
\item A sinusoidal function with an inverted rectangular pulse. This
``box'' component
adds 3 more parameters: the
duration,
midpoint, and depth
of the box, for a total of 7 parameters.
\end{enumerate}
We used MCMC fitting with 30 walkers, 15,000 steps to find the best fit with the first 3000 ``burn-in'' iterations excluded to find the best model (the corner plot for the sine+box fit is shown in Appendix \ref{AppB}, Figure \ref{Corner}). The fit to the 3.6 $\mu$m and 4.5 $\mu$m {\it Spitzer} data were done independently. We used the Bayesian Information Criterion (BIC) to determine which model is best justified
by the data. The BIC has been used
in other studies to distinguish between variable and non-variable models \citep{Naud2017, Vos2020}. 

For the 2MASS J1119-1137AB 3.6 $\mu$m {\it Spitzer} light curve, of these two models, the sine+box was most preferred. We interpret this result as evidence that the fading event
occurred. The sine+box model was preferred over the sine-only model with a $\Delta BIC = 22.2$ indicating strong evidence for a
fading event. The best fit sine+box model (blue line) is shown in the middle left panel of Figure \ref{lightcurves}. We also attempted to fit a full transit model to the event, but the data did not
provide
enough 
information to constrain the impact parameter, limb-darkening coefficients or other orbital parameters
(if the event corresponds to an exomoon transit) and the sine+box model was preferred over the sine+transit model. We extracted the Spitzer light curve using a different pipeline (described in \citealt{Vos2020}). The sine+box model was still strongly favored despite the different light curve extraction technique.

For comparison, we conducted the same analysis using the sine+box and sine fits of three other datasets: the 2MASS J1119-1137AB 4.5 $\mu$m {\it Spitzer} light curve and the WISEA J1147-2040 3.6 $\mu$m and 4.5 $\mu$m light curves. The light curves were generated using the methods described in Sch18. For these three datasets we found that the sine+box model was ruled out and the sine-only model was preferred with $\Delta BICs$ of $-4.0$, $-4.8$ and $-11.5$, respectively indicating that the anomalous event detected in the 2MASS J1119-1137AB 3.6 $\mu$m light curve is above the normal level of IPMO variability measured in the other three light curves.

Out of concern that sine+box model is not
realistic enough to model the planet's intrinsic
variability, we also fitted the data
with a Gaussian Process (GP) model, with and without a box-like dip. Our GP analysis was similar to that presented by \citet{2018MNRAS.478.4866V}. In short, we employed a quasi-periodic kernel:
\begin{equation}
\label{covar}
C_{ij} = h^2 \exp{\left[-\frac{(t_i-t_j)^2}{2\tau^2}-\Gamma \sin^2{\frac{\pi(t_i-t_j)}{T}}\right]}+\left[\sigma_i^2+\sigma_{\text{jit}}^2\right]\delta_{ij}
\end{equation}
where $C_{ij}$ is the covariance matrix; $\delta_{ij}$ is the Kronecker delta function;
$h$ is the amplitude of correlated noise; $t_i$ is the time of $i$th observation; $\tau$ is the correlation periodicity; $\Gamma$ is the ratio between the squared exponential and periodic parts of the kernel; $T$ is the period of the correlation; and
$\sigma_{\text{jit}}$ is a white noise term in addition to the reported uncertainty $\sigma_i$. We imposed Jeffreys priors, i.e., log-uniform distributions
for all parameters except $T$, for which we imposed a Gaussian prior based on the period estimated
from the sine+box function described earlier.
We adopted the following likelihood function:
\begin{equation}
\label{likelihood}
\log{\mathcal{L}} =  -\frac{N}{2}\log{2\pi}-\frac{1}{2}\log{|\bf{C}|}-\frac{1}{2}\bf{r}^{\text{T}}\bf{C} ^{-\text{1}} \bf{r}
\end{equation}
where $N$ is total number of measurements; $\bf{C}$ is the covariance matrix defined earlier; and $\bf{r}$ is the residual of observed flux minus the box transit model.

We compared the Bayesian evidence of both models (GP and GP+box) using Dynesty, a nested sampling code \citep{2020MNRAS.493.3132S}. The Bayesian evidence was then used to compute the Bayes factor (for which
$\Delta BIC$ is a proxy). We used the default settings on Dynesty. The sampling stopped automatically after standard convergence criteria were reached. We found that the GP+box model was favored by a Bayes factor of 3, which we interpret as only a marginal preference.
The GP+box model (black line) is shown in Figure \ref{lightcurves} (middle, left panel). The box parameters for both the sine+box and GP+box fits are given in Table \ref{Table:box}. Using the ratio of the depth parameter divided by its uncertainty as the SNR, the GP+box model gives a 4.2-$\sigma$ detection of an event.

\begin{deluxetable*}{cccc}[h]

\tablenum{2}
\tablecaption{Best Fit Box Parameters}
\tablewidth{0pt}
\tablehead{
\colhead{Parameter} & \colhead{Sine+box Model} & \colhead{GP+box Model}
}
\startdata
$T_{\rm mid}$ (hours) & $7.91\pm0.03$ & $7.94\pm0.06$ \\
Duration (minutes) & $38\pm3$ & $36\pm6$  \\
Depth (\%) & $0.53\pm0.05$ & $0.63\pm0.15$  \\
Depth ($R_\Earth$) & $1.59\pm0.14$ & $1.74\pm0.41$ \\
\enddata
\tablecomments{The calculation of the depth in Earth radii assumes both IPMOs have a radius of $1.38R_{\rm Jup}$ and the moon transits one of the two {\it equal flux, equal radius} (identical) planets.}
\label{Table:box}
\end{deluxetable*}

As an aside, we noted during or MCMC analysis that the joint sine+box fit led to somewhat
different rotational period of the planet then the best-fit sine curve. Table \ref{Table:sineFit} gives the best fit planet variability parameters from this work (using the sine+box fit) and Sch18. We searched for
a second periodic signal that might be attributed to
the second IPMO, but did not find any (see periodograms in Appendix \ref{AppB}, Figure \ref{periodograms}). Sch18 attempted to fit two sine-curves to the data, but were unable to identify a second rotation period. Further we note that two IPMOs with similar rotation periods and amplitudes, out-of-phase would not produce a sharp dip.

\begin{deluxetable*}{cccc}[h]
\tablenum{3}
\tablecaption{2MASS J1119-1137AB Variability Parameters}
\tablewidth{0pt}
\tablehead{
\colhead{Parameter} & \colhead{This Work$^1$} & \colhead{Sch18}
}
\startdata
Mean Flux & $1.00175\pm0.0003$ & $0.9999\pm0.0003$ \\
Amplitude (\%) & $0.17\pm0.03$ & $0.230^{+0.036}_{-0.035}$  \\
Period (hours) & $3.12^{+0.09}_{-0.08}$ & $3.02^{+0.07}_{-0.06}$  \\
Phase$^2$ (degrees) & $37\pm15$ & $29^{+16}_{-13}$ \\ 
\enddata
\centering
$^1$Based on best fit sine parameters from sine+box model.\\
$^2$Rotational Phase of the planet at t = 0.
\label{Table:sineFit}
\end{deluxetable*}

\subsection{Possible explanations for the Fading Event}

In this section, we explore the possible explanations for the fading event. The most likely source of non-astrophysical systematic noise is the IRAC detector. To check on this possibility, we searched for
(and did not find) any correlation
between the extracted aperture flux and
the centroid pixel coordinates on the {\it Spitzer}/IRAC detector. Nor did we find a significant displacement
in the centroid position at the time of the
fading event. The expected centroid shift due to a 1 hour, 1\% flux change of one of the two unresolved IPMOs is $3\times$ smaller than the 1-$\sigma$ uncertainty in the centroid position measurement on this timescale.

An eclipse of one IPMO by the other IPMO,
which is {\it a priori} very unlikely,
is ruled out by the observed
separation of 138\,mas. The change in
separation is only $\sim$2\,mas/year.
What about the possibility of an unresolved
eclipsing binary that is unrelated to the IPMOs?
Be17 found only one background star that falls
within the {\it Spitzer} aperture, at a separation of 3\farcs79.
However, with a $K$-band magnitude that is 5.7\,mag
fainter than 2MASS J1119-1137AB, even a total eclipse of this background star
would not lead to a decrease in relative flux
by the observed amount.

The two most plausible explanations are (1) the dip is caused by
the transit of a faint
object such as a moon,
disintegrating circumplanetary object, or another sort of transiently transiting debris \citep{2017ApJ...835..168D};
and (2) the dip is part of the erratic variability displayed
by the IPMO, due to clouds or other atmospheric features.
We cannot distinguish between these possibilities
with the available data.
More helpful would be
spectral observations of a fading
event, and confirmation of periodicity and
consistency of the characteristics of the fading
signals. This example highlights that the use of simultaneous multi-band or spectral observations are needed for differentiating between exomoon transits and intrinsic IPMO variability.
Because the exomoon explanation is interesting and is the topic of this paper, we will explore it a bit further, keeping in mind that erratic variability is also a reasonable explanation.

To calculate the radius of the exomoon implied by the fractional loss of light requires an estimate for the radius of the IPMO, which is not given in the literature. We used the \cite{2007ApJ...659.1661F} exoplanet models to estimate the radius of the IPMO. Specifically, we used
the models for the largest orbital distance in the library (9.5\,AU) figuring that this
was the best match to a young planet
for which the dominant heat source is internal
rather than irradiation by the star. Assuming
the objects have $25\,M_{\Earth}$ rocky cores, and accounting for uncertainties in that age and mass of the system (as reported by Be17), leads to radius
estimates of
\begin{equation}
    R_{A} = R_{B} = 1.38^{+0.17}_{-0.11}~R_{\rm Jup}.
\end{equation}
This radius is also consistent with the evolutionary models of \cite{2008ApJ...689.1327S}. Under the assumption that the IPMOs are identical,
we can calculate the moon's radius
despite the fact that we do not know which IPMO is hosting the moon.
We find $R_{\rm moon} = 1.74\pm0.19\,R_\Earth$ based on the fractional
loss of light in the GP+box model.
A more detailed characterization, including a discussion of possible orbital parameters and the likelihood of habitability, is given in Appendix \ref{AppA}.

\begin{figure}
\centering
\includegraphics[width=0.9\textwidth]{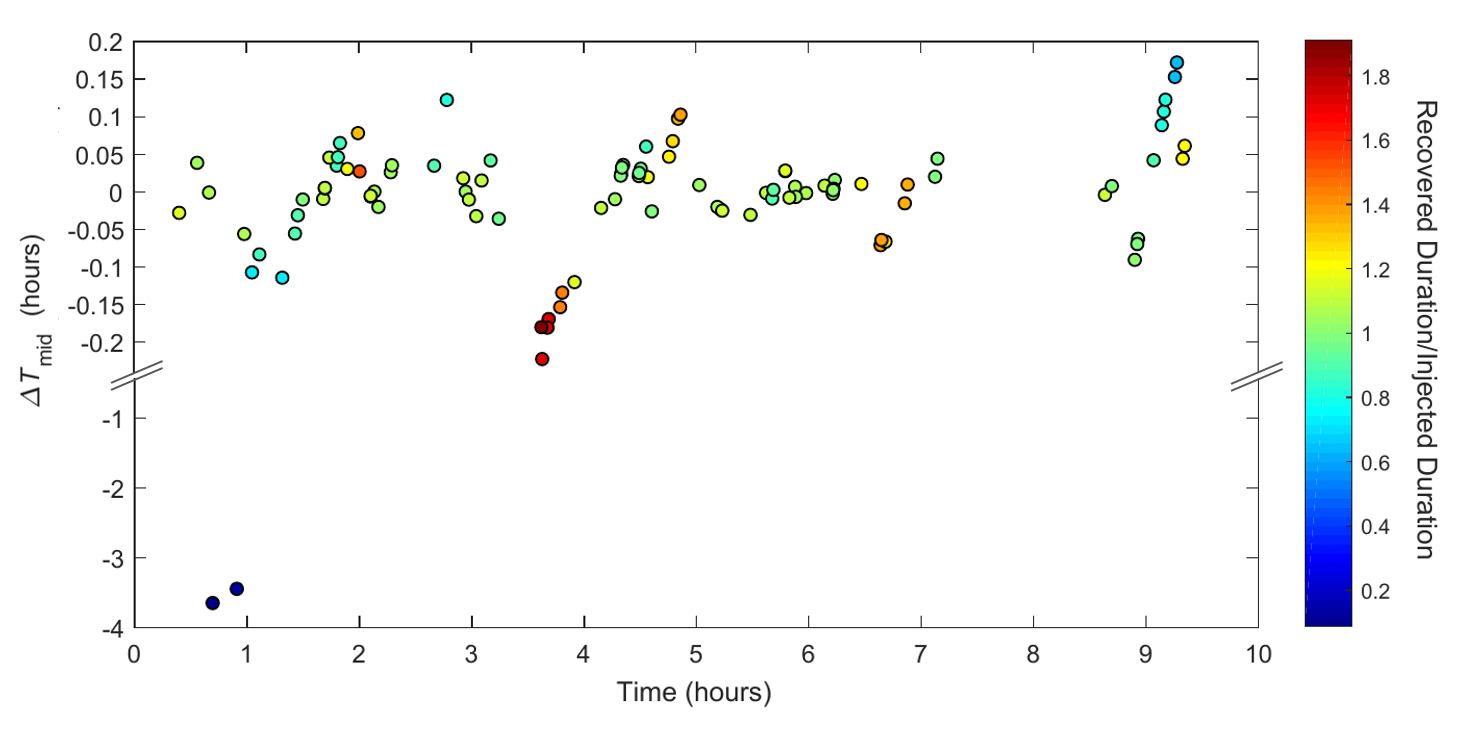}
\caption{Results of an injection/recovery test. The error in the recovered
value of the transit midpoint is
shown as a function of the time
at which the simulated transit was injected.
The color of each point indicates
the ratio of the recovered
duration to the injected duration. In 98 out of 100 injections, the correct
transit midpoint was recovered within 0.2\,hours.
In the other 2 out of 100 cases,
the sine fit was preferred over the sine+box fit indicating no detection of an event.}
\label{rainbow}
\end{figure}

To assess our sensitivity to transit signals in the data, we performed an injection/recovery test. We removed the real fading event from the data. We then injected 100
simulated transit events with the same duration and depth as the real dip, with parameters taken from the sine+box model.
We did not inject any simulated transits within the portion of the dataset covering the
real fading event.
We then fitted the simulated data with the sine and sine+box models using the same MCMC fitting method. We calculated the $\Delta BIC$ of the two models for all 100 cases, finding that the sine+box model was favored over the sine model
92\% of the time. This indicates that a transit similar to the one detected can typically be recovered in this dataset. Figure \ref{rainbow} shows the error in the recovered mid-point of transit, as a function of the time
at which the signal was injected. In all but two of the injections, the recovered transit mid-point was within 12 minutes of the input value. This test illustrates that if a 40-minute transit by a $1.7\,R_\Earth$ moon had really
occurred, we would probably have been able to detect it.

There are several possible approaches
to follow up on this candidate signal. It may be possible to follow up with an 8m-class
ground-based telescope in the K-band. A single
transit would be detectable at the $\approx$7-$\sigma$ level, and the IPMO variability cycle would be detectable after
a full night of observations. Adaptive-optics
imaging would allow the two components to be resolved; if a second transit were detected in spatially resolved observations, one could identify the host
of the exomoon. However, the 2MASS~J1119-1137AB variability has not been studied in the K-band. If the IPMOs exhibit larger-amplitude or more erratic variability at this wavelength, or if the K-band instrumental systematics are larger than anticipated, transit detection would be hindered. Further, depending on the orbital period of the exomoon, many nights of observation could be required to catch a transit.

{\it JWST} observations would be especially powerful.
From Table \ref{Table:planets} the JNSPP on this target is 66\,ppm would allow for a strong (50-$\sigma$) detection of a similar fading event even if the IPMOs were unresolved. Spectrophotometric information would allow for
the distinction between 
erratic IPMO variability (which is likely to be chromatic) and transits (for which the loss of light
should not vary strongly with wavelength).

\section{Summary}\label{sec:conc}

Based on analogies with the
properties of the large moons of the giant planets in
the solar system, the transits of moons around isolated planetary-mass
objects are expected to be (1) common, occurring for 10 to 15\% of gas giant planets, (2) frequent, with orbital periods of a few days, and (3) possible to detect, with multiple moons per system that produce transit
depths of 0.1\% to 2\%.
More than 50\% or 33 of the 57 currently
known IPMOs are sufficiently bright at near-infrared and
mid-infrared wavelengths
to allow for $>$5$-\sigma$ detection with {\it JWST} of single transits
of a Titan or Ganymede-sized moons. 

Bright, young IPMOs are favorable targets because (1) they are typically larger in radius, increasing the geometric transit probability; and (2) the brightness allows for higher precision photometry. The low masses and densities of IPMOs allow for the possibility of very close-orbiting moons with high transit probabilities. High transit probabilities are especially important because the transit search will need to be done on a target-by-target basis.

The search for exomoons transiting IPMOs will require substantial amounts of observing time on premier observatories. Will the community be willing to invest this much time? We note that thousands of hours of {\it Spitzer} and {\it Hubble Space Telescope} time was awarded to study IPMO and brown dwarf variability \citep[e.g. \textit{Weather on Other Worlds} and \textit{Cloud Atlas} programs;][]{Metchev2015, Apai2017}, independent of the motivation to search for transiting exomoons.  Furthermore, {\it JWST} is scheduled to observe the IPMO WISE 0855 in Cycle 1 for 11 hours using the NIRSpec time-series G395M spectrum \citep{2021jwst.prop.2327S}. NIRspec will achieve a spectrally-binned SNR of $\approx 4000$ in one hour\footnote{SNR was calculated with a 250K 1 $R_{\rm Jup}$ Spectrum (Sonora 2018 grid; \citealt{marley2018}) at 2.23 pc uploaded in JWST/ETC. Sensitivity is given in 1 hour and for the binned spectral sensitivity from $\lambda = 4-5\ \mu m$ ($R\approx5$).}, thus the observation will be sensitive to exomoons as small as Titan/Ganymede with 5-$\sigma$. Spectrally-resolved time-domain models are needed that will allow us to discriminate between exomoon transit events and atmospheric variations/water-ice clouds in the WISE 0855 light curve. Development of such models can be done now so that we are prepared to search for exomoons in upcoming {\it JWST} IPMO datasets.

The moons of IPMOs might be some of the most observationally
accessible habitable worlds.
For young and hot IPMOs, {\it JWST} transmission spectroscopy of moons may be possible --- and could
be easier than it would be for a planet orbiting an M dwarf in some respects.
Moons in the habitable zones of IPMOs transit every day or two,
and in some cases the transit depths may be as large as 2\%. For older and cooler IPMOs, emission spectroscopy is potentially
powerful, especially if the moon is tidally heated and comparable in temperature to its host. Further, these close-in exomoons are likely to lie in the IPMO's habitable zone (at least initially until the planet cools) which will allow us to study conditions similar to primordial earth and perhaps place limits on the time scale for the formation of life.  
If a habitable zone, 1.7$R_\Earth$ exomoon exists around of the components in 2MASS~J1119-1137AB, {\it JWST} would be capable of securely detecting a single transit.

\section*{Acknowledgements} 
We are grateful to Darren L. DePoy for many useful discussions while preparing this manuscript. We thank an anonymous referee for a helpful and thorough review.

This research has made use of the NASA Exoplanet Archive, which is operated by the California Institute of Technology, under contract with the National Aeronautics and Space Administration under the Exoplanet Exploration Program. This work has made use of the UltracoolSheet, maintained by Will Best, Trent Dupuy, Michael Liu, Rob Siverd, and Zhoujian Zhang, and developed from compilations by \cite{2012ApJS..201...19D,Dupuy_2013,2016ApJ...833...96L,Best_2017b,Best_2020b}.

{\it Software:} {\tt emcee.py} \citep{2013PASP..125..306F}, {\tt astro.py} (https://github.com/astropy/astropy), {\tt LombScargle} \citep{2017arXiv170309824V}, {\tt ELCA} \citep{2019AJ....158..243P}, {\tt Starry} \citep{Luger2019}, {\tt corner.py} \citep{corner}, {\tt numpy.py} \citep{5725236} and {\tt dynesty.py}  \citep{2020MNRAS.493.3132S}.

\appendix
\section{The 2MASS J1119-1137AB Event Interpreted as an Exomoon Transit}\label{AppA}

\begin{figure}[h]
\centering
\includegraphics[width=0.7\textwidth]{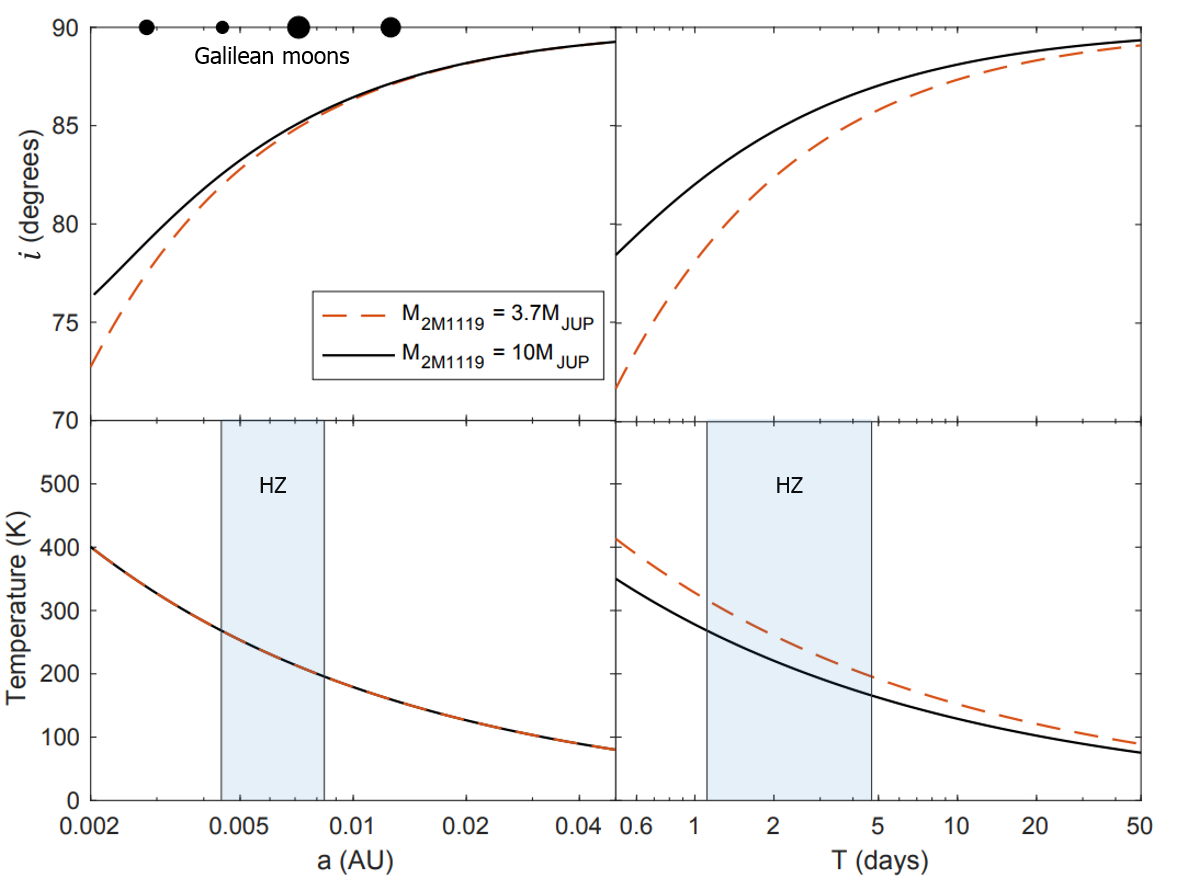}
\caption{{\bf Top panels:}~Allowed inclination of the exomoon's orbit as a function of orbital distance
(upper left) and orbital period (upper right),
for the cases $M_p = 3.7\,M_{\rm Jup}$ (blue curve)
and $10\,M_{\rm Jup}$ (red dashed curve).
The orbital distances of the four Galilean satellites are shown in the upper left, for reference.
The lower limit on the period is 0.5 days based
on the non-observation of a second transit with {\it Spitzer}.
{\bf Bottom panels:}~The equilibrium temperature of the exomoon (from Equation \ref{Teq}) as a function of orbital
distance (lower left) and period (lower right).
The shaded regions correspond to the habitable zone.}
\label{MoonPara}
\end{figure}

If the event detected in the 2MASS J1119-1137AB {\it Spitzer} light curve is due to an exomoon transit, it is possible to constrain the exomoon orbital parameter space and characteristics based on the duration of the event. If we assume the event duration corresponds to the transit duration, we can determine the distance traveled, $l$, by the exomoon during the transit assuming the IPMO is much more massive than the moon ($M_{\rm p} \gg M_{\rm m}$):
\begin{equation}\label{distTrav}
    l = T_{\rm dur}\,\sqrt\frac{GM_p}{a},
\end{equation}
where $T_{\rm dur}$ is the transit duration, $M_p$ is the mass of the IPMO, and $a$ is the exomoon's orbital distance. As
explained earlier,
the host IPMO's mass is either $M_1\approx 3.7\,M_{\rm Jup}$ or $M_2\approx 10M_{\rm Jup}$ depending on whether the system is part of the TW Hya association. To calculate the allowable parameter space for $l$, we consider both possibilities for $M_{\rm p}$,
we use the measurement
$T_{\rm dur}\approx 36$~min based on the GP fit, and allow $a$ to be a free parameter.
From $l$ we can calculate the impact parameter,
\begin{equation}\label{impact}
    b = \sqrt{1-\left(\frac{l-R_m}{2R_p}\right)^2},
\end{equation}
where $R_m$ is the radius of the moon. Using this equation for the impact parameter, we can calculate the allowable parameter space for the inclination, $i$, and the orbital period, $T$, in the usual way \citep{Winn2010}. Figure \ref{MoonPara} gives the allowed inclination of the exomoon orbit as a function of $a$ and $T$ in the upper left and right panels, respectively, for $M_{\rm p}=M_1$ (solid curve) and $M_{\rm p}=M_2$ (red dashed curve).
We can then calculate the equilibrium temperature of the exomoon with the equation

\begin{equation}\label{Teq}
T_{\rm eq} = T_{\rm p,eff}(1-\alpha_{\rm M})^{1/4}\sqrt\frac{R_{\rm p}}{2a},
\end{equation}
where $\alpha_{\rm M}$ is the exomoon's albedo and $T_{\rm p, eff}$ is the planet's effective temperature. We assume
$\alpha_{\rm M} = 0.05$ \citep{HellerBarnes_2015} and
$T_{\rm p, eff} = 1010\,K$ (Be17).
The exomoon absorbs a time and spatially-averaged flux
\begin{equation}\label{Fab}
F = 239~{\rm W/m}^2~\frac{L_{\rm p}(1-\alpha_{\rm M})}{L_\sun(1-\alpha_\Earth)} \left(\frac{1\,{\rm AU}}{a}\right)^2,
\end{equation}
where $\alpha_\Earth = 0.3$. Using $L_{\rm p} = 1.86\times 10^{-5}\,L_\sun$ (Be17) we can calculate $F$. We assume
the habitable zone for the exomoon is
defined by 83~W/m$^2 < F < 295$~W/m$^2$
(\citealt{2013ApJ...765..131K}; \citealt{Heller_2016}).  The bottom panel of Fig.~\ref{MoonPara} shows the calculated
equilibrium temperature of the exomoon (from Equation \ref{Teq}) as a function of $a$ and $T$. The shaded regions on these lower panels correspond to the habitable zone. Short-period orbits ($T \lesssim 5$ days) are more probable because (1) the transit probability of long-period orbits is low ($P <8\%$ for $T > 5$d) and (2) 20 hours of {\it Spitzer} observation covers $<17\%$ of the orbit for $T > 5$d. 
Thus, if the exomoon is real, it is quite probable that it lies in the habitable zone. The Hill radius of either IPMO in this binary system is $R_{H} = 1.3\,$AU. If the exomoon is orbiting with $T < 5$~days and $a < 0.009$~AU (approximately the same orbital distance
as Titan), then $a \ll R_{H}$ and the exomoon would be dynamically stable.

Any inference about the exomoon's mass would depend
on assumptions about its composition and on
evolutionary models, and are therefore highly
uncertain. If the exomoon has a H/He envelope, it may be much less massive than might be inferred by comparison to older objects of similar radius.  The current
sample of detected exoplanets does not contain
any planets smaller than 1.8\,$R_\oplus$ and
younger than 100\,Myr.
So, if the candidate signal correspond to an exomoon, and further follow-up confirmed the moon, not only would it be the first confirmed exomoon, it would also be the youngest terrestrial-sized object, thereby offering a unique window into the properties of terrestrial worlds in their infancy. 

\newpage
\section{Supplemental Figures from the Spitzer light curve Analysis}\label{AppB}

\begin{figure}[h]
\centering
\includegraphics[width=1\textwidth]{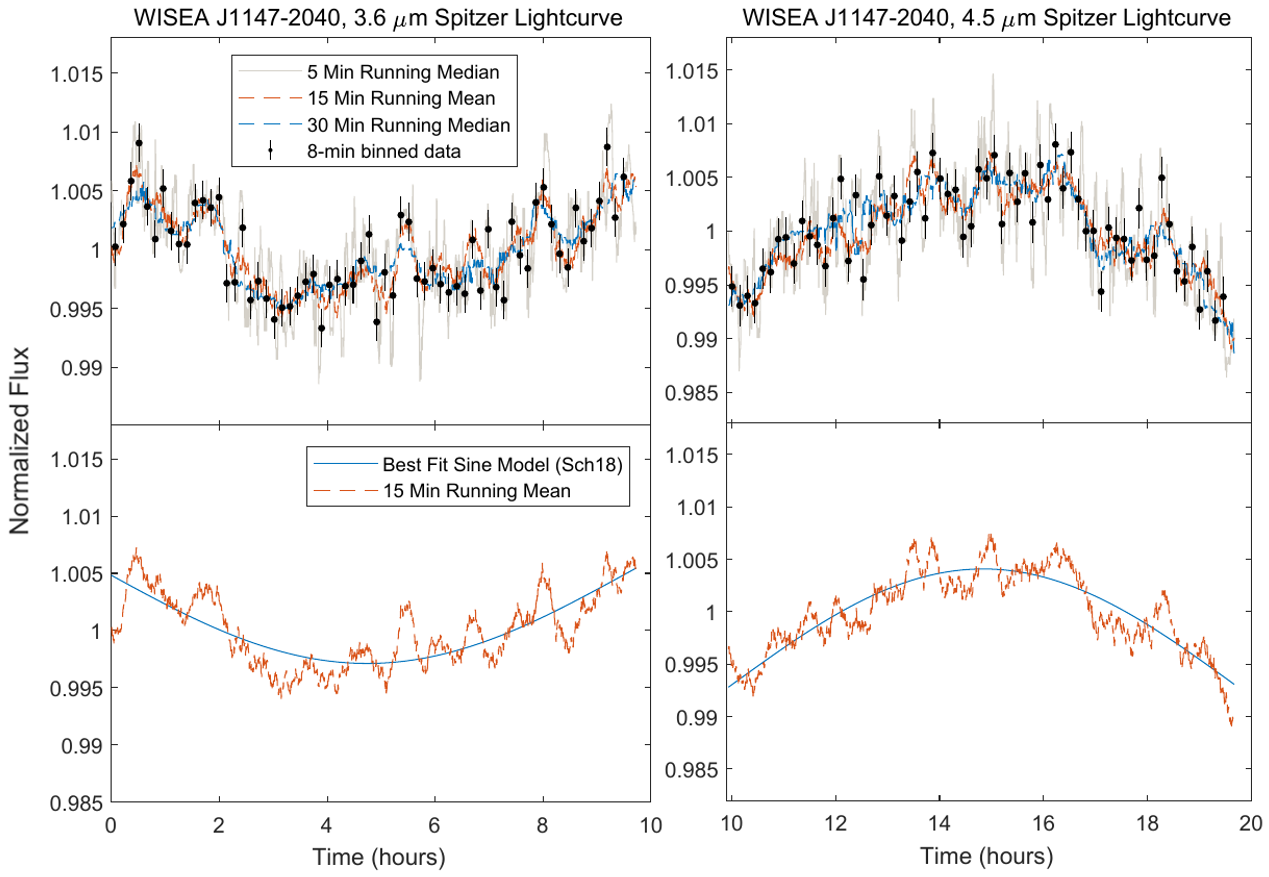}
\caption{{\bf Top:}~{\it Spitzer} light curves of WISEA J1147-2040 based on a single 20-hour observation, 10 hours
at $3.6\mu{\rm m}$ (left) and 10 hours
at $4.5\mu{\rm m}$ (right). {\bf Bottom:}~Shown in addition to the 15-min running mean of the data (red dashed line) is 
the best fit sine model (blue curves)
from the joint light curve fit given in Sch18 (since the rotation period is longer than the single-band observation, the joint fit is used in this case).}
\label{WISEAJ1147}
\end{figure}

\begin{figure}
\centering
\includegraphics[width=1\textwidth]{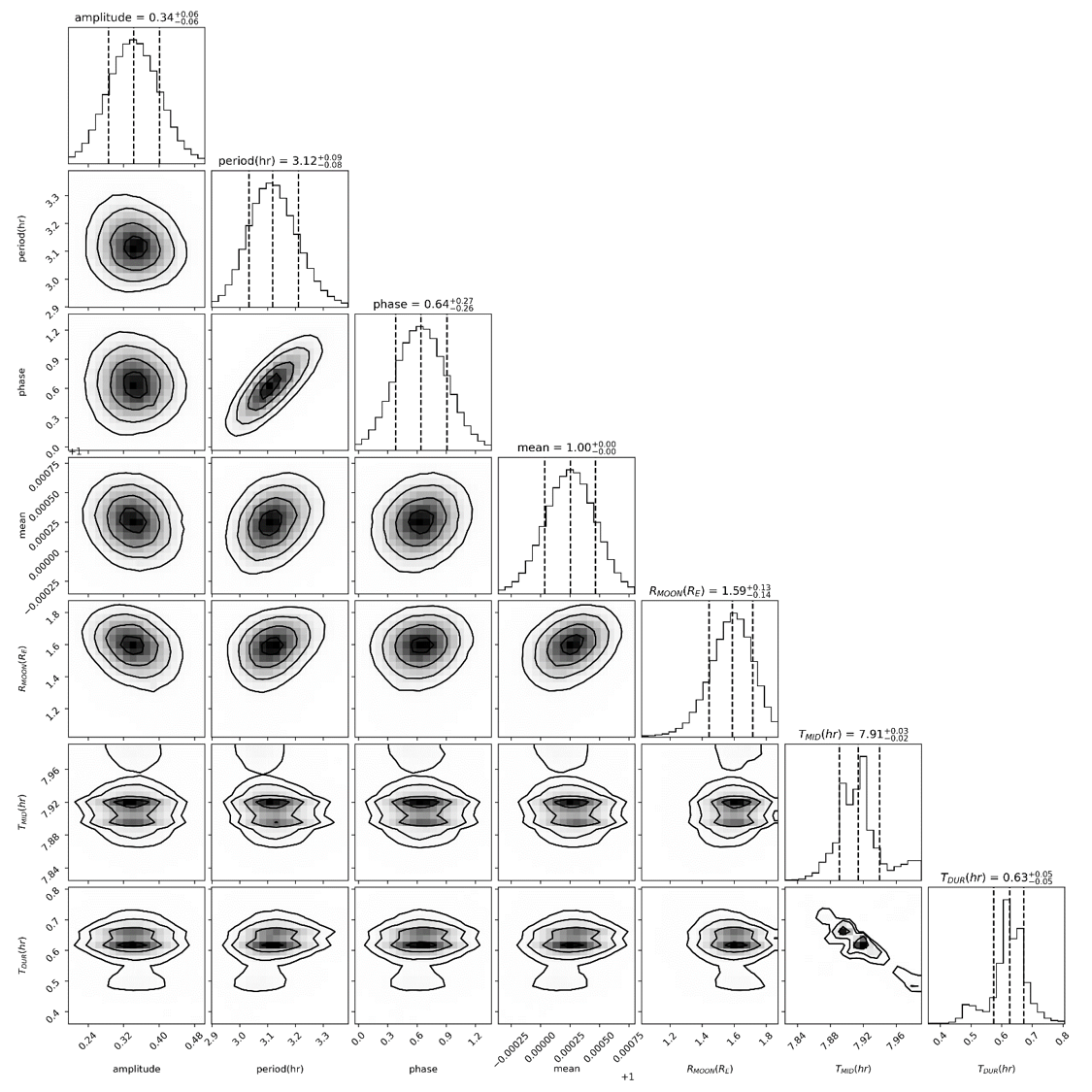}
\caption{Corner plot from the sine+box MCMC fit of the {\it Spitzer} 3.6$\mu m$ 2MASS J1119-1137AB light curve.}
\label{Corner}
\end{figure}

\begin{figure}
\centering
\includegraphics[width=1\textwidth]{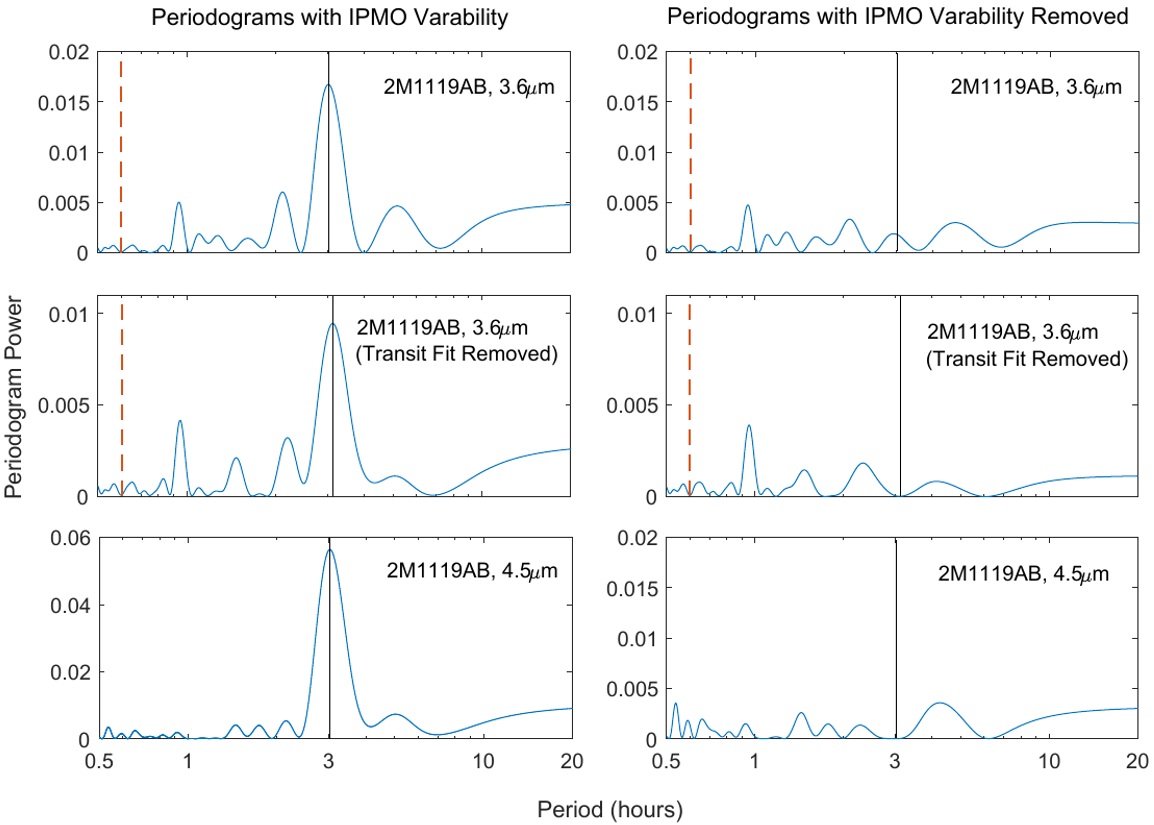}
\caption{Periodograms of 2MASS J1119-1137AB. Left panels: Periodograms include rotational varability. Right Panels: Periodograms with sinusoidal fit removed. The $3.6\mu m$ Spitzer data periodograms are shown in the upper two sets panels with the event included (top) and removed (middle). The $4.5\mu m$ Spitzer data periodograms are in the bottom panels. Red dashed line is the duration of the event. Solid line is the best-fit rotational period of the IPMO.}
\label{periodograms}
\end{figure}

\clearpage  
\bibliographystyle{aasjournal}

\end{document}